\documentclass[a4paper,12pt]{article}
\usepackage{float}
\usepackage[percent]{overpic}
\usepackage[margin=2.5cm]{geometry}
\usepackage{setspace}
\usepackage{indentfirst}
\usepackage{footmisc}
\usepackage{amsmath}
\usepackage{amssymb}
\usepackage{graphicx}
\usepackage[nosort]{cite}
\usepackage[svgnames]{xcolor}
\usepackage[colorlinks=true,
            allcolors=.,
            bookmarksnumbered=true,
            pdfpagemode=UseNone,
            pdfstartview=FitH]{hyperref}

\newcommand{\dif}{\text{d}}

\numberwithin{equation}{section}
\makeatletter
\renewcommand\section{\@startsection {section}{1}{\z@}%
	{-3.5ex \@plus -1ex \@minus -.2ex}%
	{2.3ex \@plus.2ex}%
	{\normalfont\large\sffamily\bfseries}}
\renewcommand\subsection{\@startsection{subsection}{2}{\z@}%
	{-3.25ex\@plus -1ex \@minus -.2ex}%
	{1.5ex \@plus .2ex}%
	{\normalfont\normalsize\sffamily\bfseries}}
\renewcommand\subsubsection{\@startsection{subsubsection}{3}{\z@}%
	{-3.25ex\@plus -1ex \@minus -.2ex}%
	{1.0ex \@plus .2ex}%
	{\normalfont\normalsize\sffamily\itshape}}
\makeatother

\begin{document}

\thispagestyle{empty}
\vbox{}
\vspace{0.5cm}

\begin{center}
  {\sffamily\LARGE{Dilatonic Reissner--Nordstr\"om black holes\\[2mm]
  in a Bertotti--Robinson magnetic background}}\\[10mm]
  {\sffamily\large Kenneth Hong}\\[4mm]
  {\sffamily\slshape\selectfont
    Department of Physics, National University of Singapore, Singapore
  }\\[8mm]
\end{center}
\vspace{0.7cm}

\centerline{\sffamily\bfseries Abstract}
\bigskip
\noindent
We construct an exact static, magnetically charged solution of four-dimensional Einstein--Maxwell--dilaton theory that generalizes the centered Reissner--Nordstr\"om black hole in a Bertotti--Robinson magnetic background. Its pure-background limit is the magnetic EMD deformation of Bertotti--Robinson spacetime.  Using the Emparan--Teo solution-generating technique, the charged construction reduces in canonical Weyl coordinates to replacing the Schwarzschild horizon rod by the Reissner--Nordstr\"om rod of half-length $\sigma=\sqrt{m^2-e^2}$ and integrating the resulting nonlinear background--rod interaction.  This yields a completely explicit solution for arbitrary dilaton coupling $\alpha$, with the expected special limits and numerical verification of the full field equations.

The horizon exhibits a coupled mechanical, electromagnetic and intrinsic response to the monopole--background interaction.  The residual conical stress is tied to north--south magnetic polarization; increasing the background field produces the ordered reversal scales $B_{\rm pole}<B_{\rm crit}<B_{\rm eq}$, independent of $\alpha$ at fixed seed parameters $(m,e)$, while Gauss--Bonnet identifies the conical stress with the integrated distributional Gaussian-curvature contribution at the singular pole.  Along the fixed-$(m,B,\alpha,C)$ near-extremal path considered here, every $\alpha>0$ gives vanishing horizon area and divergent curvature, whereas a correlated near-extremal strong-field limit yields a universal magnetic-reversal law.

\newpage

\section{Introduction}
\label{intro sec}

  Exact black holes in external electromagnetic fields provide useful laboratories for studying the nonlinear interaction between a compact horizon and a non-asymptotically-flat electrovacuum.  The Bertotti--Robinson (BR) spacetime is particularly distinguished: it is the homogeneous $AdS_2\times S^2$ solution of Einstein--Maxwell theory supported by a uniform electromagnetic field~\cite{Bertotti1959,Robinson1959}.  A complementary line of work began with the Harrison transformation and the Ernst construction of black holes in the Melvin magnetic universe~\cite{Harrison1968,Ernst1976}; the BR background is not an asymptotic Melvin region, but a spatially homogeneous product geometry.  Black holes in homogeneous BR-type electromagnetic environments therefore form a distinct branch.  In particular, Alekseev and Garc\'ia constructed an exact Schwarzschild black hole immersed in the homogeneous BR electromagnetic universe, whose asymptotic geometry and topology differ from the better-known Melvin magnetic universe~\cite{AlekseevGarcia1996}.  Exact Schwarzschild and rotating black holes embedded in BR-type backgrounds have received renewed attention, including ultrarelativistic and coordinate formulations adapted to their black-hole interpretation~\cite{OrtaggioAstorino2018}.

Charged and dilatonic black holes in external magnetic fields also have a substantial history in Melvin-type backgrounds.  Dilatonic C-metrics and their Ernst-type embeddings in magnetic dilaton Melvin universes were constructed for arbitrary dilaton coupling \cite{Dowker:1993bt}, while related dilatonic dihole configurations and solution-generating constructions were developed subsequently \cite{EmparanTeo2001,Liang:2001ea}.  The present problem is distinct in that the external electrovacuum is BR-type rather than Melvin-type.  Dilatonic and string-theoretic extensions of BR-type geometries have also been studied in other settings, including dyonic string backgrounds, dilaton--axion gravity, and EMD theories with nontrivial scalar potentials~\cite{LoweStrominger1994,ClementGaltsov2001,MazharimousaviEtAl2010}.

Several recent developments motivate the present work.  Ovcharenko and Podolsk\'y have developed a broad Petrov type-D description of black holes with non-aligned electromagnetic fields and identified a charged non-twisting sector describing an intrinsically charged black hole interacting with an external BR field~\cite{OvcharenkoPodolsky2026}.  Furugori and Tomizawa~\cite{FurugoriTomizawa2026} have subsequently enlarged this framework by retaining an additional independent electromagnetic parameter and clarifying several of its subclasses and limits.  Astorino~\cite{Astorino2026RNBR}, building on Alekseev's monodromy-transform construction of a charged black hole in a homogeneous BR electromagnetic field, presented the solution in a static magnetic form particularly well adapted to Weyl coordinates.  Finally, Ma, Wu and L\"u~\cite{MaWuLu2026} showed that the Schwarzschild--BR solution admits a generalization to Einstein--Maxwell--dilaton (EMD) theory.  These results naturally raise the question whether an intrinsically charged RN--BR black hole can likewise be embedded in general EMD theory.  The main result of this paper is an explicit arbitrary-$\alpha$ dilatonic deformation of the centered magnetically charged RN--BR solution, together with its horizon geometry, conical response, flux redistribution and controlled near-extremal behavior.

We answer this question using the solution-generating technique introduced by Emparan and Teo for static axisymmetric magnetic Einstein--Maxwell solutions.  Before introducing a black hole, we first apply the technique directly to the magnetic BR electrovacuum.  The result is precisely the magnetic EMD deformation of the BR background recently displayed by Ma, Wu and L\"u, after translating conventions and coordinates.  This preliminary application also makes an important conceptual distinction explicit: for nonzero dilaton coupling the Maxwell invariant remains constant, but the scalar backreaction destroys the homogeneity of the ordinary $AdS_2\times S^2$ BR geometry.  We then revisit the Schwarzschild--BR solution of Ma, Wu and L\"u and show that it too is generated directly by the same transformation.  This provides an independent derivation of their EMD solution and, crucially, exposes its auxiliary harmonic function as the sum of a finite Schwarzschild horizon rod and a separate BR-background contribution.  The nontrivial part of the charged construction is then not the transformation of the metric functions, magnetic potential and dilaton, but the replacement of the horizon-rod data and the integration of the first-order equations for $\widetilde\gamma_e$.

The central observation is geometrical.  In canonical Weyl coordinates the auxiliary harmonic function for the Schwarzschild--BR seed separates as
\begin{align}
U_{\rm SBR}=U_{\rm bg}^{\rm BR}+U_m,
\end{align}
where $U_m$ is the finite Schwarzschild horizon-rod potential and $U_{\rm bg}^{\rm BR}$ contains the BR-background contribution.  Since the RN--BR horizon rod has half-length
$\sigma=\sqrt{m^2-e^2}$, the charged construction is obtained by the replacement
\begin{align}
\label{Uansatz intro}
U_{\rm RNBR}=U_{\rm bg}^{\rm BR}+U_\sigma.
\end{align}
The transformed metric functions, magnetic potential and dilaton then follow from the Emparan--Teo transformation, while the first-order equations for $\widetilde\gamma_{\rm RNBR}$ contain a nonlinear background--rod interaction.  We integrate this interaction in closed form, obtaining a completely explicit solution that is subsequently checked numerically against the full EMD field equations.

The designation RN--BR has recently appeared in several constructions whose precise mutual identification is nontrivial.  In particular, Ovcharenko and Podolsk\'y~\cite{OvcharenkoPodolsky2026} note that the type-D Alekseev solutions are closely related to their non-aligned family, although an explicit coordinate and parameter identification has not been established.  We therefore do not assume that one recent parametrization is the unique or maximally general RN--BR representation.  Instead, our choice of seed is dictated by the requirements of the solution-generating method.  The Alekseev--Astorino representative admits a purely magnetic description and, crucially, exposes a simple canonical Weyl source structure.  It is therefore particularly well adapted to the Emparan--Teo construction, which requires an explicit magnetostatic potential.

The centered RN--BR seed used here is itself a specialization of Astorino's more general Weyl representation.  The RN horizon rod has midpoint $\beta_2$ and endpoints $\beta_2\pm\sigma$, while $\beta_1$ is a reference point associated with the BR-background Weyl sector.  Their relative axial offset is
\begin{align}
\ell=\beta_2-\beta_1.
\end{align}
The spherical solution corresponds to $\beta_1=\beta_2=0$, hence $\ell=0$.  This centered choice aligns the RN rod midpoint with the BR reference point and is the key simplification that makes the harmonic decomposition used below analytically tractable.  It is not an assumption of global equilibrium: for $Be\ne0$ the centered Einstein--Maxwell seed is generically conically unbalanced.

Beyond the construction itself, the solution provides a controlled setting in which the
interaction between intrinsic monopole charge, the external BR field and the dilaton can
be followed geometrically and electromagnetically.  We analyze the resulting conical
imbalance, the redistribution and reversal of magnetic flux across the horizon---including
the ordered scales $B_{\rm pole}<B_{\rm crit}<B_{\rm eq}$---and the associated
deformation of the intrinsic horizon geometry.  Gauss--Bonnet relates the residual
conical stress to the integrated distributional Gaussian-curvature contribution at the singular horizon pole.  We also study
the controlled near-extremal limit, which separates the temperature scaling from the
question of geometric regularity and reveals a correlated strong-field scaling of the
magnetic-reversal structure.

The paper is organized as follows.  Section~\ref{EMD sec} reviews the EMD equations and the Emparan--Teo solution-generating technique.  Sections~\ref{BR EMD sec}--\ref{SBR sec} apply it to magnetic BR and Schwarzschild--BR, isolating the background--rod decomposition that motivates the charged extension.  Section~\ref{seed sec} introduces the centered magnetic RN--BR seed and its canonical horizon rod, while Sec.~\ref{dilatonic RNBR sec} constructs and verifies the charged EMD solution.  Section~\ref{checks sec} collects its principal limits.  Section~\ref{physical properties sec} develops the nonextremal horizon physics. Section~\ref{extremal sec} then examines the near-extremal limit and correlated strong-field scaling.  We conclude in Sec.~\ref{conclusion sec}.

\section{The Emparan--Teo solution-generating technique}
\label{EMD sec}

We use the four-dimensional Einstein--Maxwell--dilaton (EMD) action:
\begin{align}
\label{EMD action}
S=\frac{1}{16\pi}\int d^4x\sqrt{-g}\left[R-2(\nabla\phi)^2-e^{-2\alpha\phi}F_{\mu\nu}F^{\mu\nu}\right].
\end{align}
Here $g_{\mu\nu}$ is the Einstein-frame spacetime metric, $R$ its Ricci scalar,
and $\phi$ the dilaton field.  The electromagnetic gauge potential and field strength are
\begin{align}
A=A_\mu\,\dif x^\mu,\qquad
F=\dif A,\qquad
F_{\mu\nu}=2\partial_{[\mu}A_{\nu]},
\end{align}
and $F^2\equiv F_{\mu\nu}F^{\mu\nu}$.  The dimensionless constant $\alpha$
controls the coupling between the dilaton and the Maxwell field. We take $\alpha\geq0$ without loss of generality, since the simultaneous replacement $(\alpha,\phi)\mapsto(-\alpha,-\phi)$ leaves the theory unchanged. Equivalently,
$e^{-2\alpha\phi}$ acts as a dilaton-dependent electromagnetic coupling.  In
the present work we restrict attention to purely magnetic, static and
axisymmetric configurations, for which
$A=A_\varphi\,\dif\varphi$.  Varying the action with respect to the metric,
gauge potential and dilaton gives, respectively, the Einstein, Maxwell and
dilaton field equations
\begin{subequations}
\label{EMD equations}
\begin{gather}
R_{\mu\nu}=2\partial_\mu\phi\partial_\nu\phi+2e^{-2\alpha\phi}\left(F_{\mu\rho}F_\nu{}^\rho-\frac14g_{\mu\nu}F^2\right),\\
\nabla_\mu\left(e^{-2\alpha\phi}F^{\mu\nu}\right)=0,\\
\nabla^2\phi+\frac{\alpha}{2}e^{-2\alpha\phi}F^2=0.
\end{gather}
\end{subequations}

For static axisymmetric configurations, EMD theory admits several related
solution-generating formulations~\cite{Yazadjiev2001}.  We use the
Emparan--Teo formulation because it acts directly on a magnetostatic
Einstein--Maxwell seed in canonical Weyl coordinates~\cite{EmparanTeo2001}.
The solution-generating technique therefore takes as input a static,
axisymmetric magnetic Einstein--Maxwell seed in the canonical Weyl form
\begin{align}
\label{Weyl seed}
ds^2=-f\,dt^2+f^{-1}\left[e^{2\gamma}(d\rho^2+dz^2)+\rho^2d\varphi^2\right],\qquad A=A_\varphi d\varphi.
\end{align}
The Weyl metric is specified by the functions $f(\rho,z)$ and
$\gamma(\rho,z)$, while $A_\varphi$ is the azimuthal component of the
magnetic potential.  The function $\gamma$ is obtained by integrating the
associated first-order equations and controls the conformal factor of the
$(\rho,z)$-sector.  The seed
functions satisfy the static axisymmetric Einstein--Maxwell equations.  We
will need explicitly the first-order equations that determine $\gamma$ once
$f$ and $A_\varphi$ are known.  In the magnetostatic sector, these equations
are
\begin{align}
\partial_\rho\gamma
&=\frac{\rho}{4f^2}\left(f_{,\rho}^2-f_{,z}^2\right)
 +\frac{f}{\rho}\left(A_{\varphi,\rho}^2-A_{\varphi,z}^2\right),\\
\partial_z\gamma
&=\frac{\rho}{2f^2}f_{,\rho}f_{,z}
 +\frac{2f}{\rho}A_{\varphi,\rho}A_{\varphi,z}.
\label{seed gamma quadrature}
\end{align}
The additive integration constant is fixed by the chosen axis normalization.

Having specified the Einstein--Maxwell seed, we now introduce
the auxiliary harmonic function required by the
Emparan--Teo solution-generating technique~\cite{EmparanTeo2001}.
The auxiliary harmonic function $\widetilde\phi$ is distinct from the seed
Weyl functions and from the physical dilaton $\phi$.  It satisfies the
axisymmetric Laplace equation
\begin{align}
\partial_\rho^2\widetilde\phi+\rho^{-1}\partial_\rho\widetilde\phi+\partial_z^2\widetilde\phi=0.
\end{align}
The transformation acts on the seed functions according to
\begin{align}
f'&=f^{1/(1+\alpha^2)}e^{-2\alpha\widetilde\phi},\\
\gamma'&=\frac{\gamma}{1+\alpha^2}+\widetilde\gamma,\\
A'_\varphi&=\frac{A_\varphi}{\sqrt{1+\alpha^2}},\\
e^{-2\phi}&=f^{\alpha/(1+\alpha^2)}e^{2\widetilde\phi}.
\end{align}
Primes distinguish the transformed metric functions and magnetic potential
from their seed counterparts.  The function $\widetilde\gamma$ obeys
\begin{align}
\partial_\rho\widetilde\gamma&=(1+\alpha^2)\rho\left[(\partial_\rho\widetilde\phi)^2-(\partial_z\widetilde\phi)^2\right],\\
\partial_z\widetilde\gamma&=2(1+\alpha^2)\rho(\partial_\rho\widetilde\phi)(\partial_z\widetilde\phi).
\end{align}
The Emparan--Teo construction leaves the choice of $\widetilde\phi$ open,
subject to the Laplace equation above.  For the applications below,
we seek the auxiliary harmonic function in the form
\begin{align}
\widetilde\phi=\frac{\alpha}{1+\alpha^2}U.
\end{align}
We take $U$ to be an axisymmetric harmonic function in
canonical Weyl coordinates, so that $\widetilde\phi$ satisfies
the required Laplace equation for every $\alpha$.
The factor $\alpha/(1+\alpha^2)$ is the normalization adopted in this
parametrization; it is not an additional field equation or a unique
consequence of the seed geometry.  With this normalization, the associated
first-order equations take the form
\begin{align}
\label{gamma U eqs}
\partial_\rho\widetilde\gamma&=\frac{\alpha^2}{1+\alpha^2}\rho(U_{,\rho}^2-U_{,z}^2),\\
\partial_z\widetilde\gamma&=\frac{2\alpha^2}{1+\alpha^2}\rho U_{,\rho}U_{,z}.
\label{gamma U z eq}
\end{align}

The remaining task is therefore to determine a suitable harmonic function
$U$ for each Einstein--Maxwell seed.  The next section applies this
seed-dependent construction to the magnetic Bertotti--Robinson solution.

\section{Dilatonic deformation of the Bertotti--Robinson background}
\label{BR EMD sec}

The magnetic Bertotti--Robinson seed provides the simplest application of
the construction in Sec.~\ref{EMD sec}.  It also supplies the background
limit against which the subsequent black-hole constructions are checked.
We begin with the Einstein--Maxwell seed and choose the auxiliary harmonic
function without assuming a previously known dilatonic solution.

The magnetic BR solution is homogeneous and locally $AdS_2\times S^2$
\cite{Bertotti1959,Robinson1959}.  In the fluxbrane coordinates $(R,u)$ used
in Ref.~\cite{MaWuLu2026}, the seed metric and magnetic potential are
\begin{align}
 ds^2_{\rm BR}&=-(1+B^2R^2)dt^2+\frac{dR^2}{1+B^2R^2}
 +\frac{4}{(1+B^2u^2)^2}\left(du^2+u^2d\varphi^2\right),
\label{BR seed metric}\\
 A_{\rm BR}&=\frac{2}{B(1+B^2u^2)}\,d\varphi,
\label{BR seed potential}
\end{align}
where an additive constant in the azimuthal component of the potential
corresponds to a local gauge choice.

Introduce local coordinates $(r,x)$, with $x=\cos\theta$, by
\begin{align}
 R=\frac{rx}{\Omega},\qquad
 u=\frac{r\sqrt{1-x^2}}{1+\Omega},
\label{BR coordinate map}
\end{align}
where
\begin{align}
 \Omega^2=1+B^2r^2(1-x^2),\qquad
 H_{\rm BR}=\frac{2\Omega}{1+\Omega}.
\label{BR H spherical}
\end{align}
The canonical Weyl coordinates are
\begin{align}
 \rho=\frac{r\sqrt{(1+B^2r^2)(1-x^2)}}{\Omega^2},\qquad
 z=\frac{rx}{\Omega^2},
\label{BR Weyl coordinates}
\end{align}
and the seed Weyl functions are
\begin{align}
 f_{\rm BR}&=\frac{1+B^2r^2}{\Omega^2},
\label{BR f spherical}\\
 e^{2\gamma_{\rm BR}}&=\frac{1+B^2r^2}{1+B^2r^2x^2}.
\label{BR gamma spherical}
\end{align}

Using the seed functions $f_{\rm BR}$ and $H_{\rm BR}$, we consider the
candidate
\begin{align}
 U_{\rm BR}=\ln\left(\frac{H_{\rm BR}}{\sqrt{f_{\rm BR}}}\right).
\label{BR U ET}
\end{align}
In the coordinates $(r,x)$, the axisymmetric Laplace equation in canonical Weyl coordinates
is equivalent to
\begin{align}
 \partial_r\!\left(\frac{r^2(1+B^2r^2)}{\Omega^2}\,
 \partial_r U_{\rm BR}\right)
 +\partial_x\!\left(\frac{1-x^2}{\Omega^2}\,
 \partial_x U_{\rm BR}\right)=0.
\label{BR harmonicity}
\end{align}
Direct substitution of Eq.~\eqref{BR U ET} verifies this equation.  Hence
\begin{align}
 \widetilde\phi_{\rm BR}=\frac{\alpha}{1+\alpha^2}U_{\rm BR}
\end{align}
is an admissible auxiliary harmonic function for the Emparan--Teo
transformation.  Integrating Eqs.~\eqref{gamma U eqs}--\eqref{gamma U z eq}
gives
\begin{align}
 \widetilde\gamma_{\rm BR}
 =\frac{\alpha^2}{1+\alpha^2}
 \left(\gamma_{\rm BR}-2\ln H_{\rm BR}\right),
\label{BR gamma tilde}
\end{align}
where the additive integration constant is set to zero.  With these
functions fixed, the transformation determines the metric, magnetic
potential and dilaton.  In particular,
\begin{align}
 e^{-2\phi}=H_{\rm BR}^{\frac{2\alpha}{1+\alpha^2}}.
\label{BR dilaton ours}
\end{align}

Returning to $(R,u)$ using Eq.~\eqref{BR coordinate map}, we have
\begin{align}
 B^2u^2=\frac{\Omega-1}{\Omega+1},\qquad
 H_{\rm BR}=1+B^2u^2\equiv H.
\label{BR H identity}
\end{align}
In the conventions of Sec.~\ref{EMD sec}, and omitting primes
from the transformed metric and magnetic potential in this
section, the solution is
\begin{align}
 ds^2={}&H^{-\frac{2\alpha^2}{1+\alpha^2}}
 \left[-(1+B^2R^2)dt^2+\frac{dR^2}{1+B^2R^2}
 +\frac{4du^2}{H^2}\right]
 +4u^2H^{-\frac{2}{1+\alpha^2}}d\varphi^2,
\label{BR EMD metric}\\
 A={}&\frac{2}{B\sqrt{1+\alpha^2}\,H}\,d\varphi,
 \qquad \phi=-\frac{\alpha}{1+\alpha^2}\ln H,
\label{BR EMD fields}
\end{align}
up to an additive gauge constant in $A_\varphi$.  The correspondence with
the conventions of Ma, Wu and L\"u~\cite{MaWuLu2026} is
\begin{align}
 \varphi_{\mathrm{MWL}}=-2\phi,\qquad
 a_{\mathrm{MWL}}=\alpha,\qquad
 \mathcal{F}_{\mathrm{MWL}}=2F.
\label{MWL conventions}
\end{align}
Under this conversion, Eqs.~\eqref{BR EMD metric} and
\eqref{BR EMD fields} reproduce their purely magnetic EMD electrovacuum:
\begin{align}
 \text{magnetic BR}\ \xrightarrow{\rm ET}\
 \text{purely magnetic EMD electrovacuum}.
\label{BR ET MWL}
\end{align}

The Maxwell invariant of this solution is constant,
\begin{align}
 F_{\mu\nu}F^{\mu\nu}=\frac{2B^2}{1+\alpha^2},
\label{BR F2 constant}
\end{align}
whereas the dilaton in Eq.~\eqref{BR EMD fields} is nonconstant for
$\alpha\ne0$ and $B\ne0$.  Constancy of the Maxwell invariant does not imply
a covariantly constant electromagnetic field or a homogeneous spacetime.
In this case, the coordinate-dependent curvature invariants reported by
Ma, Wu and L\"u~\cite{MaWuLu2026} establish that the metric is not homogeneous.

The dilaton field equation also rules out a constant dilaton in this purely
magnetic configuration when $\alpha\ne0$.  Setting $\phi=\phi_0$ would give
\begin{align}
 0=\nabla^2\phi+\frac{\alpha}{2}e^{-2\alpha\phi}F^2
 =\frac{\alpha}{2}e^{-2\alpha\phi_0}F^2,
\end{align}
which is incompatible with $\alpha\ne0$ and $F^2\ne0$.

We refer to this solution as the \emph{purely magnetic EMD electrovacuum},
or the EMD deformation of the magnetic BR background.  The homogeneous
Einstein--Maxwell BR geometry is recovered at $\alpha=0$.  The next section
applies the same construction to the Schwarzschild--BR seed.

\section{Revisiting dilatonic Schwarzschild--BR}
\label{SBR sec}

Ma, Wu and L\"u presented a dilatonic Schwarzschild--BR solution
in Einstein--Maxwell--dilaton theory~\cite{MaWuLu2026}.  In this section we
show that it follows directly from the Emparan--Teo transformation applied
to the Einstein--Maxwell Schwarzschild--BR seed.  We also decompose the auxiliary harmonic function into a Schwarzschild
rod contribution and a remainder associated with the BR background,
providing the starting point for the subsequent charged construction.

With $x=\cos\theta$, the Einstein--Maxwell seed takes the form
\begin{align}
 ds^2_{\rm SBR}=\frac{1}{\Omega^2}\left[-\frac{Q}{r^2}dt^2+\frac{r^2}{Q}dr^2+\frac{r^2}{P}\frac{dx^2}{1-x^2}+r^2P(1-x^2)d\varphi^2\right],
\label{SBR seed metric}
\end{align}
where
\begin{align}
 P&=1+B^2m^2x^2, & Q&=(1+B^2r^2)\Delta,\nonumber\\
 \Delta&=r\big[(1-B^2m^2)r-2m\big], & \Omega^2&=1+B^2r^2-B^2\Delta x^2.
\label{SBR seed functions}
\end{align}
For the static exterior considered in this section, we take
$m>0$, $B^2m^2<1$, $r>2m/(1-B^2m^2)$ and $-1<x<1$,
with $\Omega$ on the positive branch. These conditions give
$Q>0$ and $f_{\rm SBR}>0$.
The seed Maxwell field is purely magnetic.  A convenient gauge is
\begin{align}
 A_\varphi=\frac{2}{BH},\qquad
 H=\frac{2\Omega}{1+\Omega+B^2mrx^2},
\label{SBR H}
\end{align}
up to an additive constant in $A_\varphi$.  The black hole itself is uncharged; the Maxwell field belongs to the external BR electrovacuum.

The seed Weyl function is
\begin{align}
 f_{\rm SBR}=\frac{Q}{r^2\Omega^2}.
\label{SBR f definition}
\end{align}
The canonical Weyl coordinates are determined locally by
\begin{equation}
\begin{gathered}
 \rho=\frac{\sqrt{QP(1-x^2)}}{\Omega^2},\\[4pt]
 z_{,r}=-\sqrt{\frac{P(1-x^2)}{Q}}\,\rho_{,x},
 \qquad
 z_{,x}=\sqrt{\frac{Q}{P(1-x^2)}}\,\rho_{,r}.
\end{gathered}
\label{SBR Weyl coordinates}
\end{equation}
Matching the $(r,x)$ part of the metric to the canonical Weyl form then gives
\begin{align}
 e^{2\gamma_{\rm SBR}}
 =\frac{Q}{\Omega^4\left[Q\rho_{,r}^{\,2}
                  +P(1-x^2)\rho_{,x}^{\,2}\right]}.
\label{SBR gamma reconstruction}
\end{align}
We fix the additive constant in $z$ by $z(r,0)=0$, so that $z=(r-m)x$
when $B=0$.

Following Sec.~\ref{BR EMD sec}, we consider the candidate
\begin{align}
 U_{\rm SBR}=\ln\left(\frac{H}{\sqrt{f_{\rm SBR}}}\right).
\label{SBR U definition}
\end{align}
This candidate depends only on the Einstein--Maxwell seed data; the
dilatonic solution is not assumed.  Explicitly,
\begin{align}
 U_{\rm SBR}=\ln\left[\frac{2r\Omega^2}{\sqrt Q\,(1+\Omega+B^2mrx^2)}\right].
\label{SBR U explicit}
\end{align}
In the coordinates $(r,x)$, the axisymmetric Laplace equation takes
the form
\begin{align}
 \partial_r\!\left(\frac{Q}{\Omega^2}\,\partial_r U_{\rm SBR}\right)
 +\partial_x\!\left(\frac{P(1-x^2)}{\Omega^2}\,\partial_x U_{\rm SBR}\right)=0.
\label{SBR harmonicity}
\end{align}
Direct substitution of Eq.~\eqref{SBR U explicit} verifies this
equation. Hence
\begin{align}
 \widetilde\phi_{\rm SBR}=\frac{\alpha}{1+\alpha^2}U_{\rm SBR}
\end{align}
is an admissible auxiliary harmonic function for the Emparan--Teo
transformation.  

Using Eqs.~\eqref{SBR Weyl coordinates} and
\eqref{SBR gamma reconstruction} in the first-order equations
\eqref{gamma U eqs}--\eqref{gamma U z eq} gives
\begin{align}
 \widetilde\gamma_{\rm SBR}=\frac{\alpha^2}{1+\alpha^2}\left(\gamma_{\rm SBR}-2\ln H\right),
\label{SBR gamma tilde}
\end{align}
up to an additive integration constant, which we set to zero.
The Emparan--Teo transformation then yields
\begin{align}
 f'_{\rm SBR}&=f_{\rm SBR}H^{-2\alpha^2/(1+\alpha^2)},\\
 \gamma'_{\rm SBR}&=\gamma_{\rm SBR}-\frac{2\alpha^2}{1+\alpha^2}\ln H,\\
 A'_\varphi&=\frac{2}{B\sqrt{1+\alpha^2}\,H},\\
 e^{-2\phi}&=H^{2\alpha/(1+\alpha^2)}.
\label{SBR transformed functions}
\end{align}
After the convention conversion in Eq.~\eqref{MWL conventions},
the transformed metric, magnetic potential and dilaton agree with
the Schwarzschild--BR EMD solution of Ma, Wu and
L\"u~\cite{MaWuLu2026}.  This provides an independent solution-generating
derivation of their result.

For the charged extension, the key consequence is the following Weyl
decomposition.  In the zero-background limit the Schwarzschild horizon is the finite rod $-m\le z\le m$, with potential
\begin{align}
 U_m=\frac12\ln\left(\frac{R_+^{(m)}+R_-^{(m)}+2m}{R_+^{(m)}+R_-^{(m)}-2m}\right),
\qquad R_\pm^{(m)}=\sqrt{\rho^2+(z\pm m)^2}.
\end{align}
On patches where both functions are regular, $U_{\rm SBR}$
and $U_m$ satisfy the axisymmetric Laplace equation in
canonical Weyl coordinates. We therefore define
\begin{align}
 U_{\rm SBR}=U_{\rm bg}^{\rm BR}+U_m,\qquad
 U_{\rm bg}^{\rm BR}\equiv U_{\rm SBR}-U_m.
\label{SBR Weyl decomposition}
\end{align}
In the corresponding limiting Weyl coordinates,
Eq.~\eqref{SBR U explicit} gives
\begin{align}
 U_{\rm SBR}\xrightarrow{B\to0}U_m,\qquad
 U_{\rm SBR}\xrightarrow{m\to0}U_{\rm BR},
\label{SBR harmonic limits}
\end{align}
where $U_{\rm BR}$ is defined in Eq.~\eqref{BR U ET}.  Since $U_m\to0$ as
$m\to0$ away from the shrinking rod, the remainder $U_{\rm bg}^{\rm BR}$
vanishes when the external field is removed and reduces to $U_{\rm BR}$
when the black hole is removed.  These limits motivate its identification
as the background contribution to the auxiliary harmonic function.
The background contribution is compared across the different
seeds as a function of the same canonical Weyl coordinates.
The decomposition in Eq.~\eqref{SBR Weyl decomposition} provides the
starting point for the charged construction in
Sec.~\ref{dilatonic RNBR sec}.

In their broader construction, Ma, Wu and
L\"u~\cite{MaWuLu2026} introduce independent external electric
and magnetic parameters, with explicit dyonic backgrounds
and immersed Schwarzschild solutions obtained for special
couplings, including $\alpha=1$ and $\alpha=\sqrt{3}$.

The charged extension developed below instead retains the purely
magnetic external background and introduces intrinsic magnetic charge
on the black hole. The distinction is therefore between generalizing
the external electromagnetic background and charging the black hole
within that background. The constructions considered here are
summarized by
\begin{align}
 \text{BR}&\xrightarrow{\rm ET}\text{magnetic EMD electrovacuum},\nonumber\\
 \text{Schwarzschild--BR}&\xrightarrow{\rm ET}
 \text{Ma--Wu--L\"u Schwarzschild solution},\nonumber\\
 \text{RN--BR}&\xrightarrow{\rm ET}
 \text{new intrinsically charged EMD solution}.
\label{three ET levels}
\end{align}
The next section presents the magnetic RN--BR seed and its Weyl
representation.  Its dilatonic deformation is then constructed in
Sec.~\ref{dilatonic RNBR sec}.

\section{The magnetic RN--BR seed}
\label{seed sec}

Several constructions describe charged black holes in a
Bertotti--Robinson background.  Ovcharenko and
Podolsk\'y~\cite{OvcharenkoPodolsky2026} analyze a non-twisting
Petrov type-D family, including charged black holes accelerating
in the external field.  Furugori and
Tomizawa~\cite{FurugoriTomizawa2026} develop the broader
Ovcharenko--Podolsk\'y family in a parametrization retaining an
independent charge parameter.

For the present construction, we use the centered, purely magnetic
Petrov type-D RN--BR solution in the Alekseev--Astorino
representation~\cite{Astorino2026RNBR}.  Its explicit magnetic
potential and canonical Weyl representation make it suitable for
the Emparan--Teo transformation.  An explicit coordinate and parameter identification would be
required to establish its relation to particular subclasses
of these families.

\subsection{The centered magnetic seed}

Astorino's Weyl family~\cite{Astorino2026RNBR} contains two axial reference
parameters $\beta_1$ and $\beta_2$.  The horizon rod has endpoints
$\beta_2\pm\sigma$, whereas $\beta_1$ specifies an axial reference point
for the background.  A common translation changes both parameters
without changing their separation $\ell=\beta_2-\beta_1$.  We restrict
to $\ell=0$ and choose the origin so that $\beta_1=\beta_2=0$.

In coordinates $(r,x)$, $x=\cos\theta$, the seed metric can be written
\begin{align}
\label{RNBR seed}
ds^2=g_{tt}dt^2+g_{rr}\left(\frac{dr^2}{\Delta_r}+\frac{dx^2}{1-x^2}\right)-\frac{\Delta_r(1-x^2)}{g_{tt}}d\varphi^2,
\end{align}
where
\begin{align}
\Delta_r=r^2-2mr+e^2=(r-m)^2-\sigma^2,\qquad \sigma=\sqrt{m^2-e^2}.
\end{align}
We assume $m>|e|$, so that $\sigma>0$.  The parameter $e$ controls the intrinsic magnetic charge, while $B$ parametrizes the external field.
Here $g_{rr}$ multiplies the bracketed $(r,x)$ part of the metric;
the coefficient of $dr^2$ is $g_{rr}/\Delta_r$.
Throughout, the seed quantity $g_{rr}$ (and $g_{rr}$ as it reappears
in Sec.~\ref{dilatonic RNBR sec}) always denotes this bracket
coefficient, whereas the primed $g'_{rr}$ introduced after the
transformation is the coefficient of $\dif r^2$ in the generated
metric; the two differ by the factor $\Lambda/\Delta_r$.

The metric functions and magnetic potential are expressed in terms
of the auxiliary quantities
\begin{align}
x_2&=r-m,\qquad y_2=x,\qquad y_1=\frac{x_2y_2}{x_1},\\
x_1&=\frac{\sqrt2(m-r)x}{\sqrt{1-B^2\sigma^2x^2-B^2\Delta_r+\sqrt{4B^2(1+B^2\sigma^2)x^2\Delta_r+(1+B^2\sigma^2x^2-B^2\Delta_r)^2}}},
\end{align}
and
\begin{align}
\mathcal N_t&=Be\sqrt{1+B^2m^2}(mx+x_1)+(1+B^2m^2)(-m+r+my_1),\\
\mathcal D_t&=e^2+B^2e^2x_1^2-(1+B^2m^2)\Delta_r.
\end{align}
The metric functions are\footnote{We use the plus
sign in the $B^2\sigma^2x^2$ term in the denominator of $g_{rr}$, as in
Astorino's ancillary Mathematica notebook~\cite{Astorino2026RNBR}.
The printed Eq.~(3.2) has the opposite sign; direct substitution into
the Einstein--Maxwell equations confirms the ancillary-notebook sign.}
\begin{align}
g_{tt}&=-\frac{(1+B^2x_1^2)\mathcal N_t^2\Delta_r}{\mathcal D_t^2},\\
g_{rr}&=\frac{\left[Be(mx-x_1)+\sqrt{1+B^2m^2}(m-r+my_1)\right]^2}{\sqrt{4B^2(x^2-1)\Delta_r+(1+B^2\sigma^2x^2+B^2\Delta_r)^2}},
\end{align}
while the magnetic potential is
\begin{align}
A_\varphi=\frac{\left(Be x_1-\sqrt{1+B^2m^2}\,x_2\right)\left[y_2+B^2x_1(m+x_1y_2-my_2^2)\right]}{B\sqrt{1+B^2m^2}\,x_1\left\{1+B^2[x_1^2+m^2(1-y_2^2)]\right\}}+A_{\varphi0}.
\end{align}
At removable indeterminacies of this representation, including the equatorial
expressions involving $x_1$ and $y_1$, the fields are defined by continuous
extension on the chosen exterior branch.  Concretely, the square roots in
$x_1$ and in $g_{rr}$ are taken on the branch fixed by $y_1\to-1$
(equivalently $x_1\to(m-r)x$) as $B\to0$ throughout the exterior $r>r_+$.
This is the branch realized in Astorino's ancillary
notebook~\cite{Astorino2026RNBR}, and it is the branch on which the seed
reduces continuously to the magnetic Reissner--Nordstr\"om metric of
Eq.~\eqref{seed RN limit}.\footnote{The finite-$B$ expressions solve the
Einstein--Maxwell equations on either branch, but the principal square-root
branch does not commute with the $B\to0$ limit: it leaves a finite residual
(for instance, $g_{rr}$ fails to approach $r^2$).  The condition $y_1\to-1$
removes this indeterminacy.}
The constant $A_{\varphi0}$ represents the local gauge freedom and may
be chosen to make the $B\to0$ limit manifest.

Switching off the external field on this branch gives the magnetic RN seed,
\begin{align}
 B\to0:\qquad g_{tt}\to-\frac{\Delta_r}{r^2},\qquad g_{rr}\to r^2,
\label{seed RN limit}
\end{align}
with $A_\varphi$ reducing to the magnetic RN potential up to a gauge
constant.  The uncharged limit $e\to0$ belongs to the Schwarzschild--BR
family reviewed in Sec.~\ref{SBR sec}. Its identification with
that representation requires the coordinate and parameter
rescalings in Eqs.~\eqref{neutral coordinate identification}
and \eqref{neutral parameter rescaling} (the explicit map is
deferred to Sec.~\ref{checks sec}), evaluated at
$\alpha=0$, together with the corresponding azimuthal period.

\subsection{Weyl representation and horizon rod}

To apply the transformation of Sec.~\ref{EMD sec}, we write the seed in
canonical Weyl form,
\begin{equation}
ds^2=-f\,dt^2
+f^{-1}\left[e^{2\gamma}(d\rho^2+dz^2)
+\rho^2d\varphi^2\right],
\qquad A=A_\varphi\,d\varphi.
\label{RNBR seed Weyl form}
\end{equation}
The determinant of the Killing sector fixes $\rho$ through
\begin{align}
-g_{tt}g_{\varphi\varphi}=\Delta_r(1-x^2)=\rho^2.
\end{align}
The corresponding canonical Weyl coordinates are
\begin{align}
\label{Weyl coords}
\rho=\sqrt{\Delta_r(1-x^2)},\qquad z=(r-m)x.
\end{align}
They satisfy
\begin{equation}
d\rho^2+dz^2=\bigl[(r-m)^2-\sigma^2x^2\bigr]
\left(\frac{dr^2}{\Delta_r}+\frac{dx^2}{1-x^2}\right).
\label{RNBR Weyl differential identity}
\end{equation}
Comparison of Eqs.~\eqref{RNBR seed} and~\eqref{RNBR seed Weyl form}
therefore gives the seed Weyl functions
\begin{equation}
f=-g_{tt},\qquad
e^{2\gamma}=\frac{(-g_{tt})g_{rr}}{(r-m)^2-\sigma^2x^2}.
\label{RNBR seed Weyl functions}
\end{equation}
At the outer horizon $r_+=m+\sigma$,
\begin{align}
\rho=0,\qquad -\sigma\le z\le\sigma.
\end{align}
The outer horizon is therefore represented by a finite Weyl rod of
length $2\sigma$.

The centered seed is generically conically unbalanced when both the
intrinsic magnetic charge and the external field are
nonzero~\cite{Astorino2026RNBR}: the two exterior axis segments require
different azimuthal periods for regularity.  The effect of the dilatonic
deformation on the relative axis normalization is examined subsequently.
We now construct this deformation using the Emparan--Teo transformation.

\section{Dilatonic Reissner--Nordstr\"om black hole in a Bertotti--Robinson background}
\label{dilatonic RNBR sec}

We now apply the Emparan--Teo transformation to the magnetic RN--BR seed of
Sec.~\ref{seed sec}.  We choose its auxiliary harmonic function by retaining the
background contribution identified in Sec.~\ref{SBR sec} and replacing $U_m$ by
$U_\sigma$, where $\sigma=\sqrt{m^2-e^2}$.  We then integrate the first-order
equations for $\widetilde\gamma$ and obtain the metric, magnetic potential and dilaton.
Throughout this section, $f_{\rm RNBR}$, $\gamma_{\rm RNBR}$ and
$A_\varphi^{\rm RNBR}$ denote the seed functions $f$, $\gamma$ and $A_\varphi$
of Sec.~\ref{seed sec}.

We work in the nonextremal exterior, with $m>|e|$, $r>m+\sigma$ and $-1<x<1$,
on patches where the expressions below are regular.  Square-root branches are chosen by continuous connection to the zero-background
limit, consistently with the exterior branch specified in Sec.~\ref{seed sec}.

\subsection{Auxiliary harmonic function}
\label{harmonic sec}

Recall the decomposition of the Schwarzschild--BR auxiliary harmonic function in
Sec.~\ref{SBR sec},
\begin{align}
\label{SBR decomposition}
U_{\rm SBR}=U_{\rm bg}^{\rm BR}+U_m,
\end{align}
where $U_m$ is the finite-rod potential defined there.  All contributions below
are evaluated in the same canonical Weyl coordinates $(\rho,z)$.
The centered RN--BR horizon has half-length $\sigma$ rather than $m$, which motivates
the replacement
\begin{align}
\label{U RNBR}
U_{\rm RNBR}=U_{\rm bg}^{\rm BR}+U_\sigma
              =U_{\rm SBR}-U_m+U_\sigma,
\end{align}
with $R_\pm^{(a)}=\sqrt{\rho^2+(z\pm a)^2}$ and
\begin{align}
\label{U sigma}
U_\sigma=\frac12\ln\frac{R_+^{(\sigma)}+R_-^{(\sigma)}+2\sigma}
{R_+^{(\sigma)}+R_-^{(\sigma)}-2\sigma}.
\end{align}
Both contributions satisfy the axisymmetric Laplace equation
in canonical Weyl coordinates on the regular exterior patch
considered here. Moreover, $e\to0$ gives $\sigma\to m$ and $U_\sigma\to U_m$,
while $B\to0$ removes $U_{\rm bg}^{\rm BR}$ and leaves
$U_\sigma$.  These properties motivate an admissible choice of auxiliary harmonic function;
they do not fix it uniquely.  We take
\begin{align}
\label{phitilde RNBR}
\widetilde\phi_{\rm RNBR}=\frac{\alpha}{1+\alpha^2}U_{\rm RNBR},
\qquad
p\equiv\frac{\alpha^2}{1+\alpha^2}.
\end{align}
The additive constant in $U_{\rm bg}^{\rm BR}$ is retained
from the Schwarzschild--BR construction.

Define
\begin{align}
\label{K def}
K\equiv\sqrt{f_{\rm RNBR}}\,e^{U_{\rm RNBR}},
\qquad f_{\rm RNBR}=-g_{tt},
\end{align}
and introduce
\begin{align}
\label{S algebraic}
{\cal S}&=B^2\left[(r-m)^2-\sigma^2(1-x^2)\right],\\
\label{D algebraic}
{\cal D}&=\sqrt{(1-{\cal S})^2+4B^2(r-m)^2x^2},\\
\label{q algebraic}
q&=\sqrt{\frac{1-{\cal S}+{\cal D}}{2}},\\
\label{T algebraic}
T&=\frac{{\cal S}-1+{\cal D}}{2}.
\end{align}
The branch choice gives $q\to1$ as $B\to0$.  The Weyl coordinate map of
Sec.~\ref{seed sec} gives
\begin{align}
{\cal S}=B^2(\rho^2+z^2),\qquad
{\cal D}=\sqrt{(1-{\cal S})^2+4B^2z^2}.
\end{align}
Using $U_{\rm SBR}=\ln(H/\sqrt{f_{\rm SBR}})$, we write
the background contribution as
\begin{align}
U_{\rm bg}^{\rm BR}
=U_{\rm SBR}-U_m
=\ln\left(\frac{H e^{-U_m}}{\sqrt{f_{\rm SBR}}}\right)=-\ln\left(\frac{1+q}{2}\right)-\frac12\ln(1+T).
\end{align}
Here $H$ and $f_{\rm SBR}$ are the functions of
Sec.~\ref{SBR sec}, expressed in the same canonical Weyl
coordinates $(\rho,z)$ before the RN--BR coordinate map is used.
For the rod contribution,
$R_+^{(\sigma)}+R_-^{(\sigma)}=2(r-m)$ in the exterior gives
\begin{align}
U_\sigma
=\frac12\ln\frac{r-m+\sigma}{r-m-\sigma}.
\end{align}
The auxiliary quantities also obey
\begin{align}
\label{Tq identities}
T+q^2={\cal D},
\qquad
Tq^2=B^2(r-m)^2x^2.
\end{align}
Substituting the expressions for $U_{\rm bg}^{\rm BR}$ and
$U_\sigma$ into Eq.~\eqref{K def} gives
\begin{align}
\label{K explicit}
K=\sqrt{f_{\rm RNBR}}
\sqrt{\frac{r-m+\sigma}{r-m-\sigma}}
\frac{2}{(1+q)\sqrt{1+T}}.
\end{align}
The algebraic part of the Emparan--Teo transformation is therefore
\begin{align}
\label{algebraic generated fields}
f'&=f_{\rm RNBR}K^{-2p},\\
e^{-2\phi}&=K^{2\alpha/(1+\alpha^2)},\\
A'_\varphi&=\frac{A_\varphi^{\rm RNBR}}{\sqrt{1+\alpha^2}}.
\end{align}
It remains to integrate the equations for $\widetilde\gamma$ to determine
the $(r,x)$ sector of the metric.

\subsection{Determination of $\widetilde\gamma_{\rm RNBR}$ and $\Lambda$}
\label{gamma sec}

Although the auxiliary harmonic function superposes linearly,
$U_{\rm RNBR}=U_{\rm bg}^{\rm BR}+U_\sigma$, the first-order equations for $\widetilde\gamma$ are quadratic in the derivatives of $U_{\rm RNBR}$.
For two such harmonic functions $U$ and $V$, define
\begin{align}
\partial_\rho\Gamma[U,V]&=\rho\left(U_{,\rho}V_{,\rho}-U_{,z}V_{,z}\right),\\
\partial_z\Gamma[U,V]&=\rho\left(U_{,\rho}V_{,z}+U_{,z}V_{,\rho}\right).
\end{align}
We write $\Gamma_{\rm bg,bg}=\Gamma[U_{\rm bg}^{\rm BR},U_{\rm bg}^{\rm BR}]$,
$\Gamma_{{\rm bg},\sigma}=\Gamma[U_{\rm bg}^{\rm BR},U_\sigma]$ and
$\Gamma_{\sigma\sigma}=\Gamma[U_\sigma,U_\sigma]$.  The required function is
\begin{align}
\label{gamma physical decomposition}
\widetilde\gamma_{\rm RNBR}
=p\left(\Gamma_{\rm bg,bg}
+2\Gamma_{{\rm bg},\sigma}
+\Gamma_{\sigma\sigma}\right).
\end{align}
The three terms are the background, mixed and rod contributions, respectively.
Integration gives the following expressions; we adopt the displayed additive
constants, with no further constant in Eq.~\eqref{gamma physical decomposition}:
\begin{align}
\label{Gamma bgbg}
\Gamma_{\rm bg,bg}
&=\frac12\ln\left(\frac{1+T}{{\cal D}}\right)
+2\ln\left(\frac{1+q}{2}\right),\\
\label{Gamma bgsigma}
\Gamma_{{\rm bg},\sigma}
&=\frac12\ln\left[\frac{\xi+q}{\xi-q}\frac{\xi-1}{\xi+1}\right],\\
\label{Gamma sigmasigma}
\Gamma_{\sigma\sigma}
&=\frac12\ln\left(\frac{\xi^2-1}{\xi^2-x^2}\right)
=\frac12\ln\frac{\Delta_r}{\Sigma},
\end{align}
where
\begin{align}
\xi=\frac{r-m}{\sigma},
\qquad
\Sigma=(r-m)^2-\sigma^2x^2.
\end{align}
Direct differentiation verifies that these expressions satisfy
the defining first-order equations. With the additive constants
chosen above, $\Gamma_{\rm bg,bg}$ and $\Gamma_{{\rm bg},\sigma}$
vanish as $B\to0$, leaving only the rod contribution.

To express the result directly in the seed coordinates, use
\begin{align}
d\rho^2+dz^2
=\Sigma\left(\frac{dr^2}{\Delta_r}+\frac{dx^2}{1-x^2}\right),
\qquad
e^{2\gamma_{\rm RNBR}}
=\frac{f_{\rm RNBR}g_{rr}}{\Sigma},
\end{align}
and define
\begin{align}
\Lambda
=K^{2p}\exp\left(2\widetilde\gamma_{\rm RNBR}
-2p\gamma_{\rm RNBR}\right).
\end{align}
Substitution of Eqs.~\eqref{K explicit} and
\eqref{Gamma bgbg}--\eqref{Gamma sigmasigma} cancels the intermediate Weyl quantities
and yields
\begin{align}
\label{Lambda explicit}
\Lambda
=\left[
\frac{(r-m-\sigma)^2(1+q)^2}{4g_{rr}{\cal D}}
\left(\frac{r-m+\sigma q}{r-m-\sigma q}\right)^2
\right]^p.
\end{align}
This determines the $(r,x)$ sector explicitly in the seed coordinates.

\subsection{Complete solution}
\label{verification sec}

With $p=\alpha^2/(1+\alpha^2)$ and $K$, $\Lambda$ given by
Eqs.~\eqref{K explicit} and \eqref{Lambda explicit}, the metric is
\begin{align}
\label{final metric}
ds'^2={}&K^{-2p}g_{tt}dt^2
+\Lambda g_{rr}\left(\frac{dr^2}{\Delta_r}+\frac{dx^2}{1-x^2}\right)
-K^{2p}\frac{\Delta_r(1-x^2)}{g_{tt}}d\varphi^2.
\end{align}
The dilaton and magnetic potential are
\begin{align}
\label{final matter}
e^{-2\phi}=K^{2\alpha/(1+\alpha^2)},
\qquad
\phi=-\frac{\alpha}{1+\alpha^2}\ln K,
\qquad
A'=\frac{A_\varphi^{\rm RNBR}}{\sqrt{1+\alpha^2}}d\varphi.
\end{align}
The additive constant in $A_\varphi^{\rm RNBR}$ remains the usual magnetic gauge freedom.

As an independent numerical consistency check, we evaluated
the Einstein, Maxwell and dilaton equation residuals after
analytic differentiation at exterior points for several choices
of $(m,e,B,\alpha)$. The residuals vanish to numerical precision
at all sampled points.

\section{Special limits and consistency checks}
\label{checks sec}

We examine four limits of the solution constructed in
Sec.~\ref{dilatonic RNBR sec}. Unless stated otherwise,
the remaining parameters are held fixed, and the limits
are taken on regular exterior patches.
Throughout, the \emph{zero-background} limit ($B\to0$) removes the
external field while retaining the black hole, whereas the
\emph{pure-background} limit ($m,e\to0$) removes the black hole
while retaining the field; the two are distinct and are treated
separately below.

\subsection{Einstein--Maxwell limit: $\alpha\to0$}

When the dilaton coupling is removed,
\begin{align}
p\to0,\qquad K^{-2p}\to1,\qquad e^{-2\phi}\to1,\qquad A'\to A^{\rm RNBR},
\end{align}
while $\widetilde\gamma_{\rm RNBR}\to0$ and hence $\Lambda\to1$.  Therefore the complete solution reduces to the magnetic Einstein--Maxwell seed of Sec.~\ref{seed sec},
\begin{align}
(g'_{\mu\nu},A',\phi)\longrightarrow(g^{\rm RNBR}_{\mu\nu},A^{\rm RNBR},0).
\end{align}

\subsection{Uncharged black-hole limit: $e\to0$}

For vanishing intrinsic black-hole charge,
\begin{align}
\sigma\to m,\qquad U_\sigma\to U_m,\qquad
U_{\rm RNBR}\to U_{\rm SBR}.
\end{align}
The comparison with Sec.~\ref{SBR sec} also requires accounting for the
additive constants in the seed and auxiliary Weyl functions. Define
$c=1+B^2m^2$. On the common exterior branch with $B^2m^2<1$, introduce
coordinates $(\bar r,\bar x)$ by
\begin{align}
 r-m=\frac{\bar r-m}{\Omega},\qquad
 x=\frac{(1+B^2m\bar r)\bar x}{\Omega},
 \label{neutral coordinate identification}
\end{align}
where $\Omega$ is the function in Eq.~\eqref{SBR seed functions}, evaluated
at $(\bar r,\bar x;m,B)$, and $\bar r>2m/(1-B^2m^2)$. This identifies the canonical Weyl
coordinates of the two representations. In this comparison, $H$,
$\gamma_{\rm SBR}$ and $\widetilde\gamma_{\rm SBR}$ below are likewise
evaluated at $(\bar r,\bar x;m,B)$.

With the constants adopted in Secs.~\ref{seed sec} and
\ref{dilatonic RNBR sec}, direct substitution gives
\begin{align}
 \gamma_{\rm RNBR}&\longrightarrow
 \gamma_{\rm SBR}+\frac32\ln c,\nonumber\\
 \widetilde\gamma_{\rm RNBR}&\longrightarrow
 \widetilde\gamma_{\rm SBR}+p\ln c.
 \label{neutral Weyl constants}
\end{align}
Consequently,
\begin{align}
 K\longrightarrow H,\qquad
 \Lambda\longrightarrow c^{-p}H^{-2p}.
 \label{neutral K Lambda}
\end{align}
Thus, in $(t,\bar r,\bar x,\varphi)$, the uncharged metric has the same time and
azimuthal components as the Section~\ref{SBR sec} solution at $(m,B)$,
while its $(\bar r,\bar x)$ part is multiplied by $c^{3-p}$. The dilaton agrees,
and the magnetic potentials differ only by an additive gauge constant.

To recover the complete Section~\ref{SBR sec} solution, set
$\lambda=c^{(3-p)/2}$ and identify its parameters and coordinates as
\begin{align}
 \widehat m=\lambda m,\qquad
 \widehat B=\frac{B}{\lambda},\qquad
 \widehat r=\lambda\bar r,\nonumber\\
 \widehat x=\bar x,\qquad
 \widehat\varphi=\frac{\varphi}{\lambda},\qquad
 \widehat t=t.
 \label{neutral parameter rescaling}
\end{align}
This rescaling leaves $P$, $\Omega$ and $H$ unchanged and sends
$Q$ to $\lambda^2Q$. Pulling back the Section~\ref{SBR sec} metric then
multiplies its $(\bar r,\bar x)$ part by $\lambda^2=c^{3-p}$ while leaving the
time and azimuthal terms unchanged. The magnetic potential also agrees
up to a gauge constant, since
$\widehat B^{-1}\dif\widehat\varphi=B^{-1}\dif\varphi$.
The uncharged limit therefore recovers the Schwarzschild--BR EMD family
with the parameter and coordinate identification
\eqref{neutral parameter rescaling}. If $\varphi$ has period $2\pi C$,
the corresponding period of $\widehat\varphi$ is $2\pi C/\lambda$.

\subsection{Zero-background limit: $B\to0$}

Removing the external BR field gives $U_{\rm bg}^{\rm BR}\to0$.
The functions $K$ and $\Lambda$ reduce to
\begin{align}
K&\longrightarrow 1-\frac{r_-}{r},&
\Lambda&\longrightarrow
\left(1-\frac{r_-}{r}\right)^{2p},
\end{align}
where $r_\pm=m\pm\sigma$. Substitution into
Eq.~\eqref{final metric} gives
\begin{equation}
\label{EMD RN metric}
\begin{aligned}
ds'^2={}&-\left(1-\frac{r_+}{r}\right)\left(1-\frac{r_-}{r}\right)^{\frac{1-\alpha^2}{1+\alpha^2}}dt^2+\left(1-\frac{r_+}{r}\right)^{-1}\left(1-\frac{r_-}{r}\right)^{-\frac{1-\alpha^2}{1+\alpha^2}}dr^2\\
&\qquad{}+r^2\left(1-\frac{r_-}{r}\right)^{\frac{2\alpha^2}{1+\alpha^2}}\left[\frac{dx^2}{1-x^2}+(1-x^2)d\varphi^2\right].
\end{aligned}
\end{equation}
After an appropriate choice of the additive gauge
constant, the dilaton and magnetic potential are
\begin{align}
e^{-2\phi}=\left(1-\frac{r_-}{r}\right)^{\frac{2\alpha}{1+\alpha^2}},\qquad
A'=\frac{A_\varphi^{\rm RN}}{\sqrt{1+\alpha^2}}\,d\varphi.
\end{align}
This is the standard static magnetically charged EMD black hole of the Gibbons--Maeda family in the conventions of Eq.~\eqref{EMD action}; the string-coupling case $\alpha=1$ is the magnetic dual of the Garfinkle--Horowitz--Strominger solution \cite{GibbonsMaeda1988,GHS1991}.

\subsection{Pure-background limit}

We remove the black hole by taking $m,e\to0$ at fixed
$B,r,x$, along a path satisfying $m>|e|$ and away from
the shrinking rod. Then
\begin{align}
m,e\to0,\qquad \sigma\to0,\qquad U_\sigma\to0,\qquad U_{\rm RNBR}\to U_{\rm bg}^{\rm BR}.
\end{align}
The rod and mixed contributions vanish, leaving the
background contribution $\Gamma_{\rm bg,bg}$, so that
\begin{align}
 \widetilde\gamma_{\rm RNBR}\to p\,\Gamma_{\rm bg,bg}.
\end{align}
The functions $K$ and $\Lambda$ reduce to
\begin{align}
 K&\longrightarrow\frac{2}{1+q},&
 \Lambda&\longrightarrow
 \left(\frac{1+q}{2}\right)^{2p},
\end{align}
where $q$ is evaluated at $m=e=0$.
After the corresponding coordinate identification, the metric,
magnetic potential and dilaton reduce to the purely magnetic
EMD electrovacuum of Sec.~\ref{BR EMD sec}.

The uncharged and zero-background limits are summarized by
\begin{equation}
\begin{array}{ccc}
\text{dilatonic RN--BR} & \xrightarrow{\ e\to0\ } & \text{dilatonic Schwarzschild--BR}\\[1mm]
\Big\downarrow {B\to0} && \Big\downarrow {B\to0}\\[1mm]
\text{static charged EMD} & \xrightarrow{\ e\to0\ } & \text{Schwarzschild}.
\end{array}
\end{equation}

\section{Physical properties}
\label{physical properties sec}

We now turn from construction and consistency checks to the physical properties of the
nonextremal solution.  Two structures organize the discussion.  The rod half-length
$\sigma=\sqrt{m^2-e^2}$ sets the overall nonextremal horizon scale, while the coexistence
of the intrinsic magnetic monopole and the external BR field produces a north--south
interaction in the centered configuration.  We first isolate the horizon scale and its
latitude-dependent intrinsic geometry.  We then follow the interaction through three
complementary sectors: mechanically as a conical imbalance, electromagnetically as
redistribution and eventual reversal of horizon magnetic flux, and intrinsically as a
redistribution of meridional length and Gaussian curvature.  Gauss--Bonnet finally ties
the mechanical and intrinsic descriptions together by identifying the residual conical
stress with the integrated distributional Gaussian-curvature contribution at the singular horizon pole.

All horizon quantities below are obtained from the explicit fields by approaching
the outer horizon from the physical exterior:
\begin{align}
 r=r_+ +\epsilon=m+\sigma+\epsilon,\qquad \epsilon\to0^+,
 \label{horizon expansion}
\end{align}
All square-root branches are fixed by continuation from the physical exterior.
This prescription separates local horizon quantities from those depending on the
azimuthal period chosen to regularize an exterior-axis segment.

\subsection{Horizon scale and geometry}
\label{horizon geometry thermodynamics sec}

The outer nonextremal Killing horizon is
\begin{align}
 r_H=r_+=m+\sigma,\qquad \sigma=\sqrt{m^2-e^2}.
 \label{horizon location}
\end{align}
Near the outer horizon the complete metric has the leading behavior
\begin{align}
 -g'_{tt}=a(x)\epsilon+O(\epsilon^2),\qquad
 g'_{rr}=\frac{b(x)}{\epsilon}+O(1),
\label{horizon leading metric}
\end{align}

Here $g'_{rr}$ is the coefficient of $\dif r^2$ in the transformed
metric, whereas the seed quantity $g_{rr}$ in Sec.~\ref{seed sec}
multiplies the bracketed $(r,x)$ part. The two angular components remain
finite at interior latitudes. Although they depend on $x$, their determinant
does not. We retain $p=\alpha^2/(1+\alpha^2)$ as defined in
Sec.~\ref{dilatonic RNBR sec}.
For the induced metric $h_H$ on a spatial cross-section of the horizon,
with coordinates $(x,\varphi)$ evaluated at $r=r_H$, the area element is
\[
 dA_H
 =\sqrt{\det h_H}\,dx\,d\varphi
 \equiv {\cal H}_H\,dx\,d\varphi .
\]

The corresponding horizon area density is independent of $x$ and evaluates to
\begin{align}
 {
 {\cal H}_H
 =
 \frac{
 4^p\sigma^{2p}(m+\sigma)^{2/(1+\alpha^2)}
 }{
 (1+B^2\sigma^2)^p
 (1+B^2m^2)^{1/[2(1+\alpha^2)]}
 }}.
 \label{horizon area density}
\end{align}
Keeping the azimuthal identification general,
\begin{align}
 0\leq\varphi<2\pi C,
\end{align}
the horizon area is therefore
\begin{align}
 {A_H=4\pi C\,{\cal H}_H}.
 \label{horizon area}
\end{align}

\subsubsection{Surface gravity and temperature}

We compute the surface gravity with respect to the Killing vector
$\partial_t$, using the time coordinate inherited from the seed metric.
At fixed horizon latitude, introduce the proper radial distance $\ell$
in the $(t,r)$ sector. The near-horizon metric then assumes the local
Rindler form
\begin{equation}
 ds^2_{(t,r)}\simeq -\kappa^2\ell^2dt^2+d\ell^2,
\end{equation}
and hence the surface gravity is determined by the ratio of the coefficients in
Eq.~\eqref{horizon leading metric}:
\begin{align}
 \kappa=\frac12\sqrt{\frac{a(x)}{b(x)}}.
\end{align}

The separate coefficients $a(x)$ and $b(x)$ depend on the horizon latitude,
but their ratio does not.  Substitution yields
\begin{align}
 {
 \kappa=
 \frac{
 \sigma^{(1-\alpha^2)/(1+\alpha^2)}
 (1+B^2m^2)^{1/[2(1+\alpha^2)]}
 (1+B^2\sigma^2)^{\alpha^2/(1+\alpha^2)}
 }{
 4^{\alpha^2/(1+\alpha^2)}
 (m+\sigma)^{2/(1+\alpha^2)}
 }}.
 \label{surface gravity}
\end{align}
Euclidean regularity then gives the Hawking temperature
\begin{align}
 {T_H=\frac{\kappa}{2\pi}}.
 \label{Hawking temperature}
\end{align}
Combining Eqs.~\eqref{horizon area density} and \eqref{surface gravity} gives
\begin{align}
 {\kappa{\cal H}_H=\sigma},
 \qquad
 \kappa A_H=4\pi C\,\sigma.
 \label{kappa area identity}
\end{align}
For fixed azimuthal normalization $C$, the explicit $B$- and $\alpha$-dependence
cancels from this product.
Thus $\kappa{\cal H}_H$ measures the Weyl rod half-length $\sigma$,
even though the horizon area density and surface gravity separately depend
on the background field and dilaton coupling.

\subsubsection{Induced horizon metric}

The induced horizon metric takes the constant-determinant form
\begin{align}
 {
 d\ell_H^2={\cal H}_H
 \left[\frac{dx^2}{F(x)}+F(x)\,d\varphi^2\right]},
 \label{horizon canonical metric}
\end{align}
The bracketed metric has unit determinant in these coordinates, so that
\[
 \det h_H={\cal H}_H^{\,2},
 \qquad
 dA_H={\cal H}_H\,dx\,d\varphi .
\]
Here $F(x)$ is a dimensionless metric function, distinct from the Maxwell
two-form $F=\dif A$. The shape function is
\begin{align}
 F(x)={}&(1-x^2)
 (1+B^2\sigma^2)^p
 (1+B^2m^2)^{1/[2(1+\alpha^2)]}
 \nonumber\\
 &\times
 \frac{
 (1+B^2\sigma^2x^2)^{(1-\alpha^2)/(1+\alpha^2)}
 }{
 \left[1+B^2m^2-Be\sqrt{1+B^2m^2}\,x\right]^{2/(1+\alpha^2)}
 }.
 \label{horizon shape function}
\end{align}
In the coordinates $(x,\varphi)$, the area density is independent of latitude.
The constant-determinant form separates the overall scale ${\cal H}_H$ from
the angular geometry encoded by $F(x)$; it does not imply an intrinsically
homogeneous horizon. For $Be\ne0$, $F(x)$ is not symmetric under $x\to-x$,
leading to unequal axis regularity conditions and north--south deformation.
The magnetic-flux
distribution is determined separately by the Maxwell field and will be
compared with these geometrical quantities below.

\subsection{Conical response and mechanical imbalance}
\label{conical response sec}

Conical defects and their removal by external electromagnetic fields are familiar in C-metric, Ernst and black-dihole geometries.  In the dilatonic C-metric, for example, a magnetic background can supply the force required to remove nodal singularities, while dilatonic dihole constructions exhibit related balance and unbalanced-charge phenomena \cite{Dowker:1993bt,EmparanTeo2001,Liang:2001ea}.  The centered RN--BR configuration considered here behaves differently: when both the intrinsic monopole charge and the BR field are nonzero, one exterior-axis segment necessarily retains a conical defect.
For period $2\pi C$, regularity at a simple pole requires
$C|F'(x_{\rm pole})|=2$.  The normalizations that regularize the north and south poles
separately are
\begin{align}
 C_N={}&
 \frac{
 \left[1+B^2m^2-Be\sqrt{1+B^2m^2}\right]^{2/(1+\alpha^2)}
 }{
 (1+B^2\sigma^2)^{1/(1+\alpha^2)}
 (1+B^2m^2)^{1/[2(1+\alpha^2)]}
 },
 \label{CN horizon}\\
 C_S={}&
 \frac{
 \left[1+B^2m^2+Be\sqrt{1+B^2m^2}\right]^{2/(1+\alpha^2)}
 }{
 (1+B^2\sigma^2)^{1/(1+\alpha^2)}
 (1+B^2m^2)^{1/[2(1+\alpha^2)]}
 }.
 \label{CS horizon}
\end{align}
Consequently
\begin{align}
 {
 \frac{C_N}{C_S}
 =
 \left[
 \frac{1+B^2m^2-Be\sqrt{1+B^2m^2}}
      {1+B^2m^2+Be\sqrt{1+B^2m^2}}
 \right]^{2/(1+\alpha^2)}}.
 \label{axis ratio horizon}
\end{align}
For finite real $\alpha$, simultaneous regularity therefore requires $Be=0$; the
centered charged solution in a nonzero BR field is generically conically unbalanced.

The same normalizations apply along the connected exterior-axis segments.
At fixed exterior radius $r$, let $s_\perp$ be proper distance from the
axis in the transverse two-surface and let $\mathcal L_\varphi$ be the
circumference of an azimuthal orbit. Expanding the full metric at
$x=1-\eta$ or $x=-1+\eta$, with $\eta\to0^+$, gives
\begin{align}
 \lim_{s_\perp\to0}\frac{\mathcal L_\varphi}{s_\perp}
 =\begin{cases}
 2\pi C/C_N,&x\to1,\\
 2\pi C/C_S,&x\to-1.
 \end{cases}
 \label{axis radial constancy}
\end{align}
These limits are independent of $r$ on each connected exterior-axis segment.
Thus $C_N$ and $C_S$ in Eqs.~\eqref{CN horizon}--\eqref{CS horizon} regularize the
entire connected north and south exterior-axis segments, respectively, rather than only
the corresponding horizon poles.  With the convention
$0\leq\varphi<2\pi C$, one may choose either $C=C_N$ or $C=C_S$; the other exterior
axis then retains a conical singularity.

The residual conical singularity measures the mechanical imbalance of
the centered black-hole/background configuration. Its tension is a global
conical stress associated with the azimuthal identification, rather than
a local Lorentz force or a thermodynamic interaction energy.

Consider first the north-regular completion $C=C_N$.  The south exterior axis then
has deficit angle
\begin{align}
 \delta_S
 =2\pi\left(1-\frac{C_N}{C_S}\right),
\end{align}
and we introduce the standard dimensionless conical-tension parameter
\begin{align}
 \mu_S\equiv\frac{\delta_S}{8\pi}
 =\frac14\left(1-\frac{C_N}{C_S}\right).
 \label{muS definition}
\end{align}
For the present solution it is useful to define
\begin{align}
 Y\equiv\frac{Be}{\sqrt{1+B^2m^2}},
 \qquad |Y|<1
 \label{interaction Y}
\end{align}
in the nonextremal sector $|e|<m$.  The exact axis ratio becomes
\begin{align}
 \frac{C_N}{C_S}
 =
 \left(\frac{1-Y}{1+Y}\right)^{2/(1+\alpha^2)},
\end{align}
so that
\begin{align}
 \mu_S(Y,\alpha)
 =
 \frac14\left[
 1-
 \left(\frac{1-Y}{1+Y}\right)^{2/(1+\alpha^2)}
 \right].
 \label{exact conical tension}
\end{align}
All dependence on $(B,m,e)$ in this axis ratio and residual tension
enters through $Y$. In particular, the conical imbalance vanishes if
either $B=0$ or $e=0$.
For $Be>0$, one has $0<Y<1$ and hence
\begin{align}
 0<\mu_S<\frac14,\qquad 0<\delta_S<2\pi.
\end{align}
The residual south axis is then deficit- or string-type.
The alternative
south-regular completion $C=C_S$ leaves a north-axis excess. Defining
$\mu_N=\tfrac14(1-C_S/C_N)$ for that choice, the tensions satisfy
\begin{align}
 \mu_N(Y,\alpha)=\mu_S(-Y,\alpha).
\end{align}
This is consistent with the north--south interchange under $B\to-B$.

The weak-field expansion begins as
\begin{align}
 \mu_S
 =
 \frac{eB}{1+\alpha^2}
 -
 \frac{2e^2B^2}{(1+\alpha^2)^2}
 +O(B^3).
 \label{mu weak B}
\end{align}
Consequently the linear conical susceptibility is
\begin{align}
 \chi_S
 \equiv
 \left.\frac{d\mu_S}{dB}\right|_{B=0}
 =
 \frac{e}{1+\alpha^2}.
 \label{mu susceptibility}
\end{align}
The product $Be$ controls the leading mechanical imbalance, while the
dilaton coupling suppresses the linear susceptibility. At $\alpha=0$, the weak-field response begins as
$\mu_S=eB+O(B^2)$. For fixed $e>0$, increasing $\alpha^2$ reduces
the linear susceptibility. The negative quadratic coefficient implies
an initially sublinear response for $eB>0$; it does not by itself
establish a separate screening mechanism.

The exact expression also determines the nonlinear strong-field behavior. For $e>0$,
\begin{align}
 \frac{dY}{dB}
 =
 \frac{e}{\left(1+B^2m^2\right)^{3/2}}
 >0 ,
 \label{Y monotonic}
\end{align}
so $\mu_S$ increases monotonically with the external field. Nevertheless the response
remains bounded: at fixed $e/m<1$,
\begin{align}
 Y&\longrightarrow \frac{e}{m},\\
 \mu_S&\longrightarrow
 \mu_\infty
 =
 \frac14
 \left[
 1-
 \left(
 \frac{m-e}{m+e}
 \right)^{2/(1+\alpha^2)}
 \right],
 \qquad B\to\infty .
 \label{mu infinity}
\end{align}
Thus the centered geometry approaches a finite limiting imbalance rather than an
unbounded conical response. The exact expression in Eq.~\eqref{exact conical tension}
also shows that increasing $\alpha^2$ suppresses the positive tension
at fixed $0<Y<1$. We now turn to the spatially resolved electromagnetic
manifestation of the same monopole--background interaction.

\subsection{Magnetic-flux redistribution and reversal}
\label{magnetic redistribution sec}

We now examine the magnetic field on the horizon, distinguishing the
net monopole flux from its distribution between latitudes. In this
subsection, $A$ denotes the transformed magnetic potential $A'$ of
Sec.~\ref{dilatonic RNBR sec}, and $F=\dif A$ denotes its field strength.

Magnetic flux through black-hole horizons has long been used to characterize the interaction between horizons and external electromagnetic fields, both in the test-field approximation and in exact magnetized geometries.  Hemispheric and horizon-cap fluxes exhibit nonlinear strong-field behavior and, in appropriate extremal limits, the black-hole Meissner effect \cite{BicakJanis1985,KarasVokrouhlicky1991,KarasBudinova2000,GibbonsPangPope2014}.  Here the intrinsic monopole charge makes both the net flux and its hemispheric distribution relevant.

\subsubsection{Magnetic charge and hemispheric fluxes}

The magnetic charge enclosed by a closed constant-$t,r$ two-surface
$\mathcal{S}$ is defined by
\begin{align}
P=\frac{1}{4\pi}\int_{\mathcal{S}}F .
\label{magnetic charge definition}
\end{align}
For the purely magnetic gauge potential
\begin{align}
A=A_\varphi(r,x)\,d\varphi ,
\end{align}
the pullback of the field strength to $\mathcal{S}$ is
\begin{align}
F\big|_{\mathcal{S}}
=\partial_xA_\varphi\,dx\wedge d\varphi .
\end{align}
We orient the enclosing two-surfaces by $\dif x\wedge\dif\varphi$
and use the corresponding sign convention for magnetic charge and normal
flux density. With $-1\leq x\leq1$ and $0\leq\varphi<2\pi C$, one finds
\begin{align}
P
&=\frac{1}{4\pi}\int_0^{2\pi C}d\varphi
\int_{-1}^{1}dx\,\partial_xA_\varphi
\nonumber\\
&=\frac{C}{2}\left[A_\varphi(r,+1)-A_\varphi(r,-1)\right]
\nonumber\\
&=\frac{Ce}{\sqrt{1+\alpha^2}\sqrt{1+B^2m^2}} .
\label{magnetic charge}
\end{align}
The result is independent of $r$, as follows from $dF=0$: the flux of
$F$ is unchanged between closed constant-$r$ two-surfaces enclosing
the black hole. It is also insensitive to an additive gauge constant
in $A_\varphi$. Thus the seed parameter $e$ is proportional to, but is
not by itself identical with, the physical magnetic charge $P$ when the
azimuthal normalization, dilaton coupling, and background field are
nontrivial. The factor $C$ arises from the azimuthal period
$0\leq\varphi<2\pi C$.

This conservation statement concerns enclosing
surfaces within a given solution. Along a family with fixed $(m,e,\alpha)$
and varying $B$, the magnetic charge need not remain fixed, particularly
when $C=C_N(B)$ is imposed.

To resolve how the total magnetic flux is distributed over the horizon, divide
the horizon into southern and northern hemispheres. Their oriented magnetic
fluxes are
\begin{align}
\Phi_S
&=\int_{\mathcal{H}_S}F
=2\pi C\left[A_{\varphi,H}(0)-A_{\varphi,H}(-1)\right],
\label{south flux}\\
\Phi_N
&=\int_{\mathcal{H}_N}F
=2\pi C\left[A_{\varphi,H}(+1)-A_{\varphi,H}(0)\right].
\label{north flux}
\end{align}
Their sum gives the total magnetic flux,
\begin{align}
\Phi_N+\Phi_S=4\pi P .
\end{align}
For later convenience define
\begin{align}
\Delta A_S&\equiv A_{\varphi,H}(0)-A_{\varphi,H}(-1),\\
\Delta A_N&\equiv A_{\varphi,H}(+1)-A_{\varphi,H}(0),
\end{align}
so that
\begin{align}
\Phi_S=2\pi C\,\Delta A_S,\qquad
\Phi_N=2\pi C\,\Delta A_N .
\end{align}
It is useful to record the sum and north--south difference,
\begin{align}
 \Delta A_\Sigma
 &\equiv
 \Delta A_N+\Delta A_S
 =
 \frac{2e}
 {\sqrt{1+\alpha^2}\sqrt{1+B^2m^2}},
 \label{flux sum}\\
 \Delta A_{\rm NS}
 &\equiv
 \Delta A_N-\Delta A_S
 =
 \frac{2Bm(m+\sigma)}
 {(1+B^2m^2)\sqrt{1+\alpha^2}}.
 \label{flux NS}
\end{align}
Here $\Delta A_\Sigma$ is the net-flux combination and satisfies
$2\pi C\,\Delta A_\Sigma=4\pi P$.  By contrast,
$\Delta A_{\rm NS}$ is only the north--south oriented-flux difference; it is not a
second conserved charge and should not in general be identified with a unique
``external-field flux.''  In the uncharged case $e=0$, however,
$\Delta A_\Sigma=0$ while $\Delta A_N=-\Delta A_S$.  With $B>0$ directed along the
positive symmetry axis, the normal field is outward in the north and inward in the south:
BR magnetic flux enters through the southern hemisphere and leaves through the northern
hemisphere.

We use magnetic polarization to describe the north--south difference
in the oriented horizon fluxes. It is not an additional conserved charge
or an asymptotic magnetic dipole moment. The uncharged example shows
that nonzero hemispheric fluxes can coexist with vanishing total flux.

Equivalently,
\begin{align}
 \Delta A_N&=
 \frac{1}{\sqrt{1+\alpha^2}}
 \left[
 \frac{e}{\sqrt{1+B^2m^2}}
 +\frac{Bm(m+\sigma)}{1+B^2m^2}
 \right],\\
 \Delta A_S&=
 \frac{1}{\sqrt{1+\alpha^2}}
 \left[
 \frac{e}{\sqrt{1+B^2m^2}}
 -\frac{Bm(m+\sigma)}{1+B^2m^2}
 \right].
\end{align}
For $e\ne0$, define
\begin{align}
 Z\equiv\frac{\Delta A_{\rm NS}}{\Delta A_\Sigma}
 =
 \frac{Bm(m+\sigma)}{e\sqrt{1+B^2m^2}}.
\end{align}
The oriented fractions are $f_N=(1+Z)/2$ and $f_S=(1-Z)/2$.
These oriented fractions can lie outside $[0,1]$ when one of the
integrated hemispheric fluxes changes sign.

For $Be>0$, the oriented south-hemisphere flux changes sign when
\begin{align}
 e\sqrt{1+B^2m^2}=Bm(m+\sigma),
\end{align}
or equivalently
\begin{align}
 B_{\rm crit}^2=\frac{m-\sigma}{2m^2\sigma}.
 \label{hemisphere critical B}
\end{align}
For the opposite sign of $Be$, the corresponding statement is reflected
north--south.

The magnetic redistribution is also directly tied to the conical interaction
variable.  From Eqs.~\eqref{interaction Y} and the definition of \(Z\),
\begin{align}
 Z=\frac{m}{m-\sigma}\,Y.
 \label{ZY relation}
\end{align}
For fixed $e/m$, $Y$ parametrizes the oriented flux polarization through
Eq.~\eqref{ZY relation}, while together with $\alpha$ it determines the residual
conical tension through Eq.~\eqref{exact conical tension}. The value $Z=1$ marks the vanishing
of the integrated southern flux for $e>0$ and $B>0$.

The preceding relations are global: they describe the conserved monopole charge and the
oriented redistribution of its flux between the two hemispheres.  We now resolve the same
redistribution locally on the horizon.

\subsubsection{Local magnetic field and reversal}

\label{local magnetic field sec}

The integrated hemispheric fluxes admit a direct local interpretation.  On the
physical exterior branch, evaluating the explicit fields at the horizon gives
\begin{align}
 x_{1,H}=-\sigma x,\qquad y_{1,H}=-1,
\end{align}
and, up to an irrelevant additive gauge constant,
\begin{align}
 A_{\varphi,H}(x)
 =
 -\frac{
 [1+B^2\sigma((m+\sigma)x^2-m)]
 [Be\,x+\sqrt{1+B^2m^2}]
 }{
 B\sqrt{1+B^2m^2}\sqrt{1+\alpha^2}
 [-1+B^2(e^2x^2-m^2)]
 }.
 \label{horizon magnetic potential}
\end{align}
The local normal magnetic flux density is
\begin{align}
 {\cal B}_H(x)
 \equiv
 \frac{F_{x\varphi,H}}{\sqrt{\det h_H}}
 =
 \frac{\partial_xA_{\varphi,H}}{{\cal H}_H}.
 \label{local horizon magnetic field}
\end{align}
Since ${\cal H}_H>0$ and is independent of $x$,
\begin{align}
 \operatorname{sgn}{\cal B}_H(x)
 =
 \operatorname{sgn}F_{x\varphi,H}(x).
 \label{magnetic field sign equivalence}
\end{align}
Consequently the zeros of $F_{x\varphi,H}$ are precisely the local
magnetic-field reversal latitudes.

Integrating this gauge-invariant density over the horizon and its two
hemispheres gives Eqs.~\eqref{magnetic charge}--\eqref{north flux}.  In the
zero-background limit,
\begin{align}
 F_{x\varphi,H}\xrightarrow[B\to0]{}
 \frac{e}{\sqrt{1+\alpha^2}},
\end{align}
which is the uniform coordinate component of the isolated magnetic EMD
field on the horizon. Its normal flux density is obtained by dividing by
${\cal H}_H$ evaluated at $B=0$.

For the reversal analysis, take $0<e<m$ and $B\geq0$; the other sign
choices follow by reversing the field orientation and reflecting the
horizon north--south. At fixed seed parameters $(m,e,B)$, the zeros of
$F_{x\varphi,H}$ are independent of $\alpha$: the expression
$F_{x\varphi,H}=\partial_xA_{\varphi,H}$ has an overall positive factor
$1/\sqrt{1+\alpha^2}$. The normal field magnitude and the horizon geometry nevertheless depend on
$\alpha$.

The reversal equation simplifies by cancelling a positive factor in the
horizon potential. Using $Y$ from Eq.~\eqref{interaction Y}, define
\begin{align}
 c_0=1-B^2m\sigma,\qquad c_2=B^2\sigma(m+\sigma).
\end{align}
Since
$1+B^2(m^2-e^2x^2)=(1+B^2m^2)(1-Yx)(1+Yx)$,
Eq.~\eqref{horizon magnetic potential} becomes, for $B>0$,
\begin{align}
 A_{\varphi,H}
 =\frac{c_0+c_2x^2}
 {B\sqrt{1+\alpha^2}(1+B^2m^2)(1-Yx)},
 \label{horizon potential reduced}
\end{align}
up to an additive gauge constant. Thus
\begin{align}
 F_{x\varphi,H}
 &=\frac{\mathcal Q_H(x)}
 {B\sqrt{1+\alpha^2}(1+B^2m^2)(1-Yx)^2},\\
 \mathcal Q_H(x)&=Yc_0+2c_2x-Yc_2x^2.
 \label{reversal quadratic}
\end{align}
All denominator factors are positive on $-1\leq x\leq1$. Moreover,
\begin{align}
 \mathcal Q_H'(x)=2c_2(1-Yx)>0,
\end{align}
so there can be at most one local reversal latitude.
At the north pole,
\begin{align}
 \mathcal Q_H(1)
 =Y+B^2\sigma\bigl[2(m+\sigma)-Y(2m+\sigma)\bigr]>0.
\end{align}
Consequently any reversed region is a connected south-polar cap.
The
zero that can enter the horizon interval has the explicit form
\begin{align}
 x_0(B)
 =\frac{\sqrt{1+B^2m^2}
 -\sqrt{m/\sigma}\sqrt{1+B^2\sigma^2}}{Be}.
 \label{exact reversal latitude}
\end{align}
It moves strictly northward as $B$ increases, since
\begin{align}
 \frac{dx_0}{dB}
 =\frac{1}{eB^2}\left[
 \frac{\sqrt{m/\sigma}}{\sqrt{1+B^2\sigma^2}}
 -\frac{1}{\sqrt{1+B^2m^2}}\right]>0.
 \label{reversal latitude monotonic}
\end{align}
Here $m>\sigma>0$ makes the bracket positive.
As $B\to0^+$,
$x_0\to-\infty$; the root first becomes physical when it reaches $x=-1$.
At $B=0$, the field is positive everywhere for $e>0$.

Three thresholds distinguish local reversal from integrated-flux
reversal. The condition $F_{x\varphi,H}(-1)=0$ gives the south-pole
threshold
\begin{align}
 B_{\rm pole}^2=
 \frac{m-\sigma}
 {\sigma(4m^2+3m\sigma+\sigma^2)}.
 \label{Bpole exact}
\end{align}
The integrated threshold $B_{\rm crit}$ is defined by $\Phi_S=0$ in
Eq.~\eqref{hemisphere critical B}. The local reversal boundary reaches
the equator when
\begin{align}
 F_{x\varphi,H}(0)=0,\qquad
 B_{\rm eq}^2=\frac{1}{m\sigma}.
 \label{Beq}
\end{align}
The ratios
\begin{align}
 \frac{B_{\rm pole}^2}{B_{\rm crit}^2}
 &=\frac{2m^2}{4m^2+3m\sigma+\sigma^2}<1,\\
 \frac{B_{\rm crit}^2}{B_{\rm eq}^2}
 &=\frac{m-\sigma}{2m}<1
\end{align}
establish the nonextremal hierarchy
\begin{align}
 B_{\rm pole}<B_{\rm crit}<B_{\rm eq}.
 \label{magnetic threshold hierarchy}
\end{align}
For $0<e/m<1$, these scales mark the appearance of a reversed south-polar
cap, the change of sign of the integrated southern flux, and the passage
of the local zero through the equator. In particular, a reversed cap
exists while $\Phi_S$ remains positive when
$B_{\rm pole}<B<B_{\rm crit}$. The thresholds are independent of $\alpha$
at fixed $(m,e)$; this statement does not assume that the physical charge
$P$ is held fixed as the parameters vary.

The limiting cases help interpret this magnetic response.  The isolated magnetic EMD black hole carries
an outward intrinsic monopole field whose normal flux density is independent of latitude.
By contrast, the Schwarzschild--BR limit has no intrinsic magnetic charge, but the BR
field still threads the horizon.  For \(B>0\) directed along the positive symmetry axis,
the normal BR field is outward in the north and inward in the south, so flux enters through
the southern hemisphere and leaves through the northern one while the net closed-surface
flux vanishes.

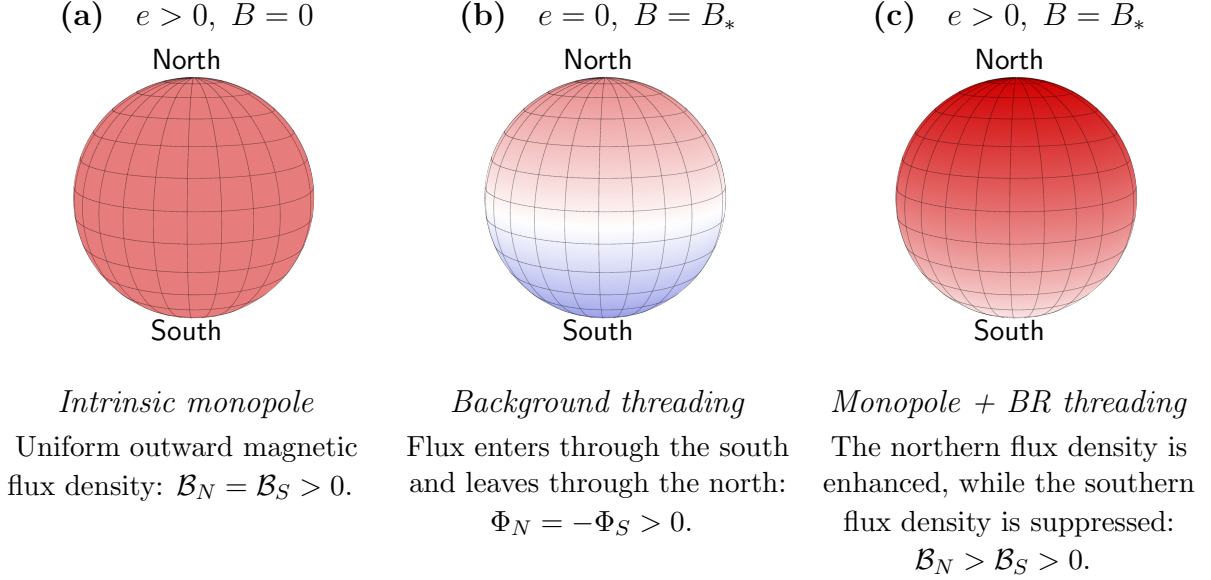
\begin{figure}[H]
\centering
\begin{minipage}[t]{0.32\textwidth}
\centering
\textbf{(a)}\quad $e>0,\;B=0$

\vspace{1mm}
\begin{overpic}[width=\linewidth]{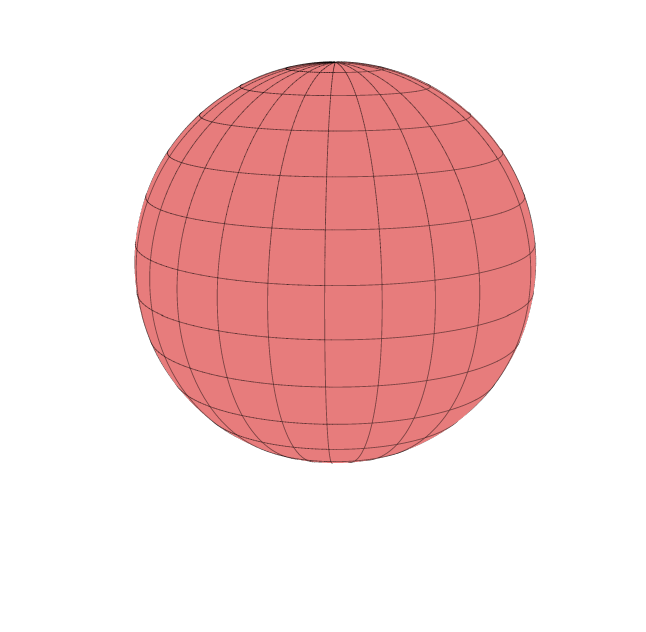}
\put(50,93){\makebox(0,0){\small\textsf{North}}}
\put(50,23){\makebox(0,0){\small\textsf{South}}}
\end{overpic}

\vspace{-6mm}
\begin{minipage}[t]{\linewidth}
\centering
\textit{Intrinsic monopole}

\vspace{1mm}
{\small
Uniform outward magnetic\\
flux density: \({\cal B}_N={\cal B}_S>0\).
}
\end{minipage}
\end{minipage}
\hfill
\begin{minipage}[t]{0.32\textwidth}
\centering
\textbf{(b)}\quad $e=0,\;B=B_*$

\vspace{1mm}
\begin{overpic}[width=\linewidth]{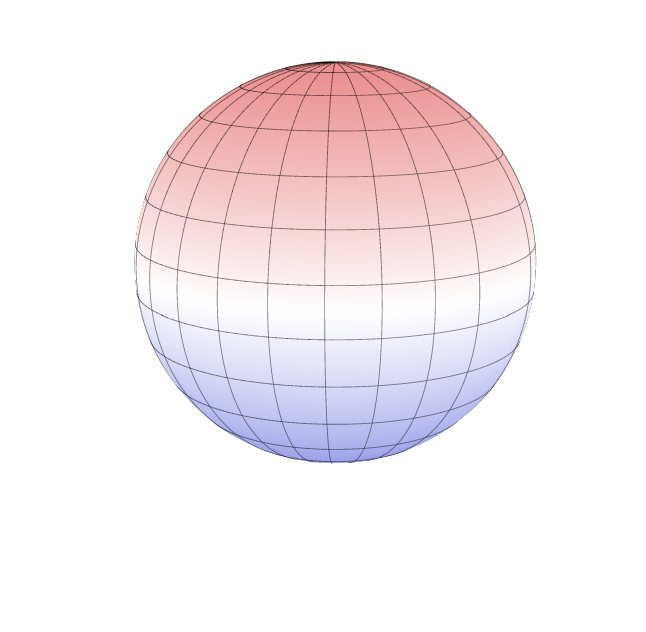}
\put(50,93){\makebox(0,0){\small\textsf{North}}}
\put(50,23){\makebox(0,0){\small\textsf{South}}}
\end{overpic}

\vspace{-6mm}
\begin{minipage}[t]{\linewidth}
\centering
\textit{Background threading}

\vspace{1mm}
{\small
Flux enters through the south\\
and leaves through the north:\\
\(\Phi_N=-\Phi_S>0\).
}
\end{minipage}
\end{minipage}
\hfill
\begin{minipage}[t]{0.32\textwidth}
\centering
\textbf{(c)}\quad $e>0,\;B=B_*$

\vspace{1mm}
\begin{overpic}[width=\linewidth]{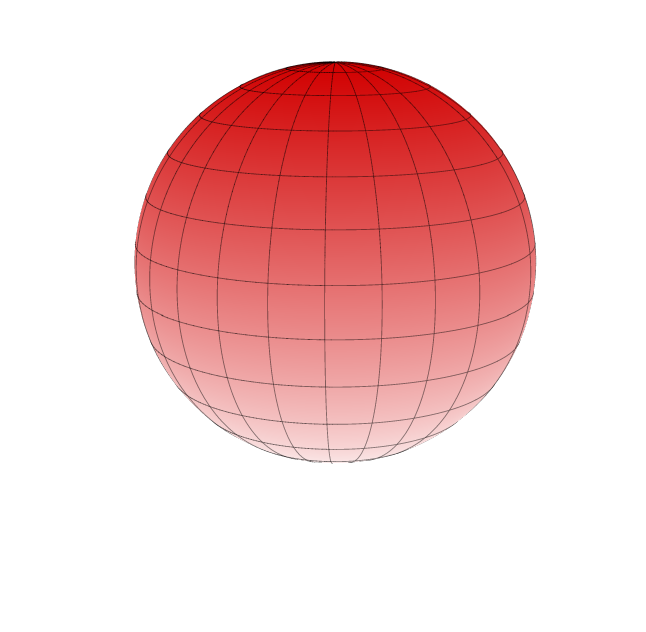}
\put(50,93){\makebox(0,0){\small\textsf{North}}}
\put(50,23){\makebox(0,0){\small\textsf{South}}}
\end{overpic}

\vspace{-6mm}
\begin{minipage}[t]{\linewidth}
\centering
\textit{Monopole + BR threading}

\vspace{1mm}
{\small
The northern flux density is\\
enhanced, while the southern\\
flux density is suppressed: \({\cal B}_N>{\cal B}_S>0\).
}
\end{minipage}
\end{minipage}

\vspace{4mm}
\caption{Comparison of horizon magnetic-field configurations.  Panel (a) shows the
uniform intrinsic monopole flux of the isolated magnetic EMD black hole.  Panel (b)
isolates BR threading in the Schwarzschild--BR limit: the horizon is threaded even
though the net magnetic charge vanishes.  Panel (c) combines the intrinsic monopole
with the same BR field used in panel (b), producing a north--south redistribution
without local field reversal.  Here \(B_*=0.9B_{\rm pole}\); the parameters
\(m=2\) and \(\alpha=4/5\) are common, while panels (a) and (c) use
\(e=7/10\). The spherical surfaces display the angular field distribution;
they are not isometric embeddings of the horizon.}
\label{horizon field origin fig}
\end{figure}

The charged solution combines these two effects nonlinearly: the BR
threading reinforces the outward monopole field in the north and opposes it in the south.
For the pole values used in Fig.~\ref{horizon field origin fig}, define
\begin{equation*}
 {\cal B}_N\equiv{\cal B}_H(1),\qquad
 {\cal B}_S\equiv{\cal B}_H(-1).
\end{equation*}
Figure~\ref{horizon field origin fig} compares the normal field
${\cal B}_H$ in the isolated, uncharged immersed, and charged immersed
solutions. Panels (b) and (c) have the same seed values of $(m,B,\alpha)$,
with $B_*=0.9B_{\rm pole}$ evaluated for the charged member. Their
comparison includes the change in geometry associated with adding charge;
it is not an additive decomposition of the nonlinear solution.

The local change relative to the isolated solution can be quantified by
subtracting its uniform monopole baseline. For $e\ne0$, define the fractional change in the local normal magnetic-flux density
\begin{align}
 \Pi_H(x;B)
 \equiv
 \frac{{\cal B}_H(x;B)-{\cal B}_H(x;0)}
 {{\cal B}_H(x;0)} .
 \label{fractional horizon flux redistribution}
\end{align}
This comparison holds $(m,e,\alpha)$ fixed and includes changes in both the
Maxwell field and the horizon area density. In particular,
\begin{equation*}
 1+\Pi_H(x;B)
 =\frac{F_{x\varphi,H}(x;B)}{e/\sqrt{1+\alpha^2}}
  \frac{{\cal H}_H(0)}{{\cal H}_H(B)}.
\end{equation*}
It does not compare solutions at fixed physical magnetic charge.
For the $e>0$ family, $\Pi_H>0$ denotes enhancement relative to the
isolated magnetic black hole and $\Pi_H<0$ denotes suppression.
The local reversal boundary is
\begin{align}
 {\cal B}_H=0\quad\Longleftrightarrow\quad\Pi_H=-1,
\end{align}
whereas $\Pi_H=0$ means equality with the isolated reference field.

\begin{figure}[!tbp]
\centering
\begin{minipage}[t]{0.47\textwidth}
\centering
\textbf{(a)}\quad $B=0$

\vspace{1mm}
\includegraphics[width=\linewidth,trim=0 25 0 25,clip]{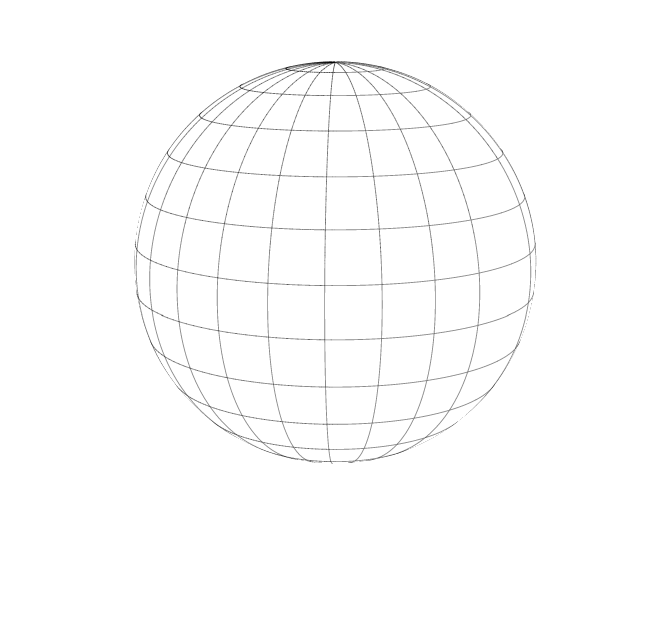}

\vspace{-8mm}
\textit{Reference monopole}

{\small No redistribution: \(\Pi_H=0\).}
\end{minipage}
\hfill
\begin{minipage}[t]{0.47\textwidth}
\centering
\textbf{(b)}\quad $0<B<B_{\rm pole}$

\vspace{1mm}
\includegraphics[width=\linewidth,trim=0 25 0 25,clip]{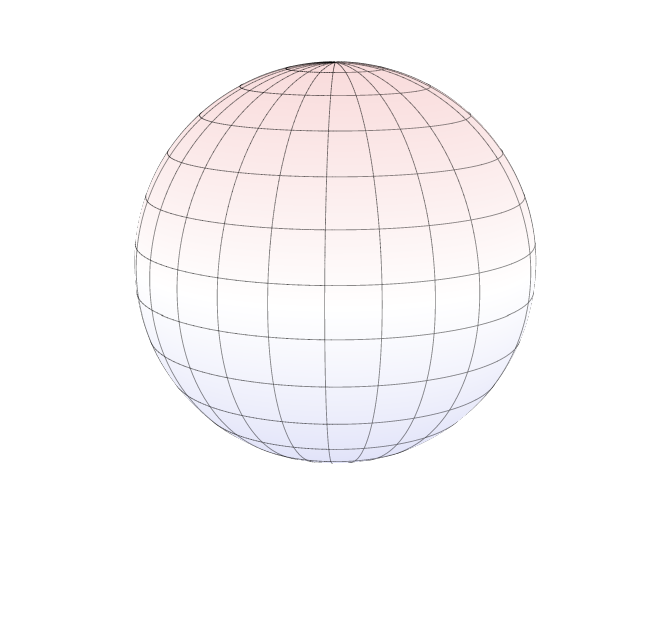}

\vspace{-8mm}
\textit{North--south redistribution}

{\small Northern normal flux density is enhanced and southern density suppressed,\\
but \({\cal B}_H>0\) everywhere.}
\end{minipage}

\par\vspace{10mm}\noindent
\begin{minipage}[t]{0.47\textwidth}
\centering
\textbf{(c)}\quad $B_{\rm pole}<B<B_{\rm crit}$

\vspace{1mm}
\includegraphics[width=\linewidth,trim=0 25 0 25,clip]{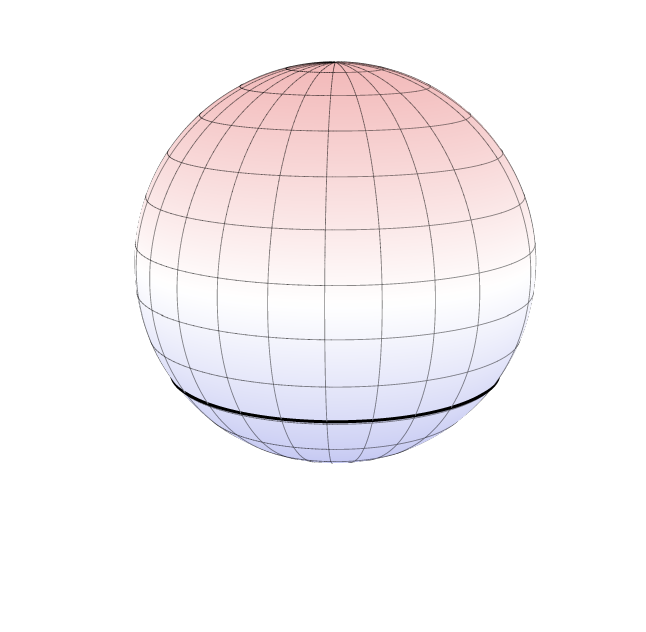}

\vspace{-8mm}
\textit{Local reversal}

{\small A reversed south-polar cap has\\
formed, but the integrated southern\\
flux still satisfies \(\Phi_S>0\).}
\end{minipage}
\hfill
\begin{minipage}[t]{0.47\textwidth}
\centering
\textbf{(d)}\quad $B_{\rm crit}<B<B_{\rm eq}$

\vspace{1mm}
\includegraphics[width=\linewidth,trim=0 25 0 25,clip]{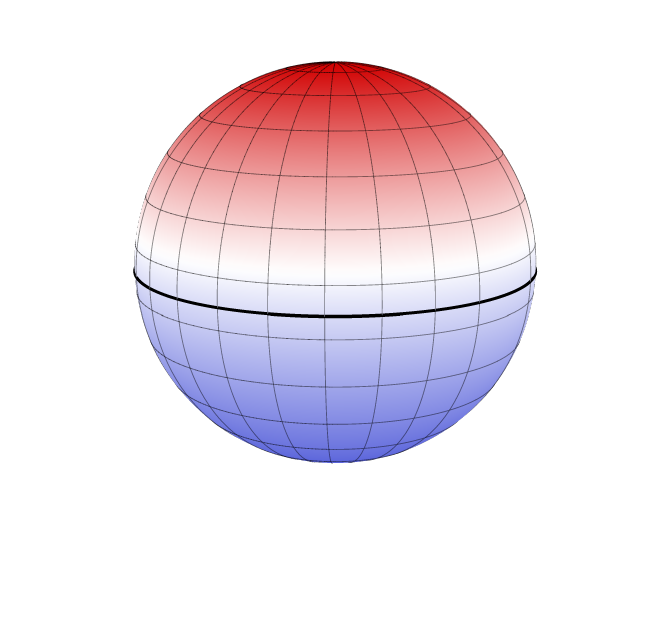}

\vspace{-8mm}
\textit{Hemispheric reversal}

{\small The reversed region is large enough that the\\
integrated southern flux satisfies \(\Phi_S<0\).}
\end{minipage}

\vspace{4mm}
\caption{Evolution of the horizon magnetic-flux distribution with increasing BR
field. Colors show the fractional change in normal flux density
\(\Pi_H=[{\cal B}_H(B)-{\cal B}_H(0)]/{\cal B}_H(0)\): red denotes enhancement
and blue suppression relative to the isolated magnetic black hole.  The black
curve marks the local reversal boundary \({\cal B}_H=0\), not
\(\Pi_H=0\).  Representative field strengths are chosen within the indicated
regimes; their numerical values are not part of the classification. The
spherical surfaces display the angular field distribution and are not
isometric embeddings of the horizon.}
\label{horizon flux redistribution fig}
\end{figure}

Figure~\ref{horizon flux redistribution fig} illustrates the reversal
sequence for $(m,e,\alpha)=(2,7/10,4/5)$. The colors show $\Pi_H$,
and black curves mark ${\cal B}_H=0$. For this family,
$B_{\rm pole}\simeq0.0469$, $B_{\rm crit}\simeq0.0919$, and
$B_{\rm eq}\simeq0.5166$. At $B=B_{\rm crit}$, the local zero has
already moved to $x_0\simeq-0.4923$: the vanishing southern flux
results from cancellation between negative flux through the
reversed south-polar cap and positive flux through the remainder
of the southern hemisphere.

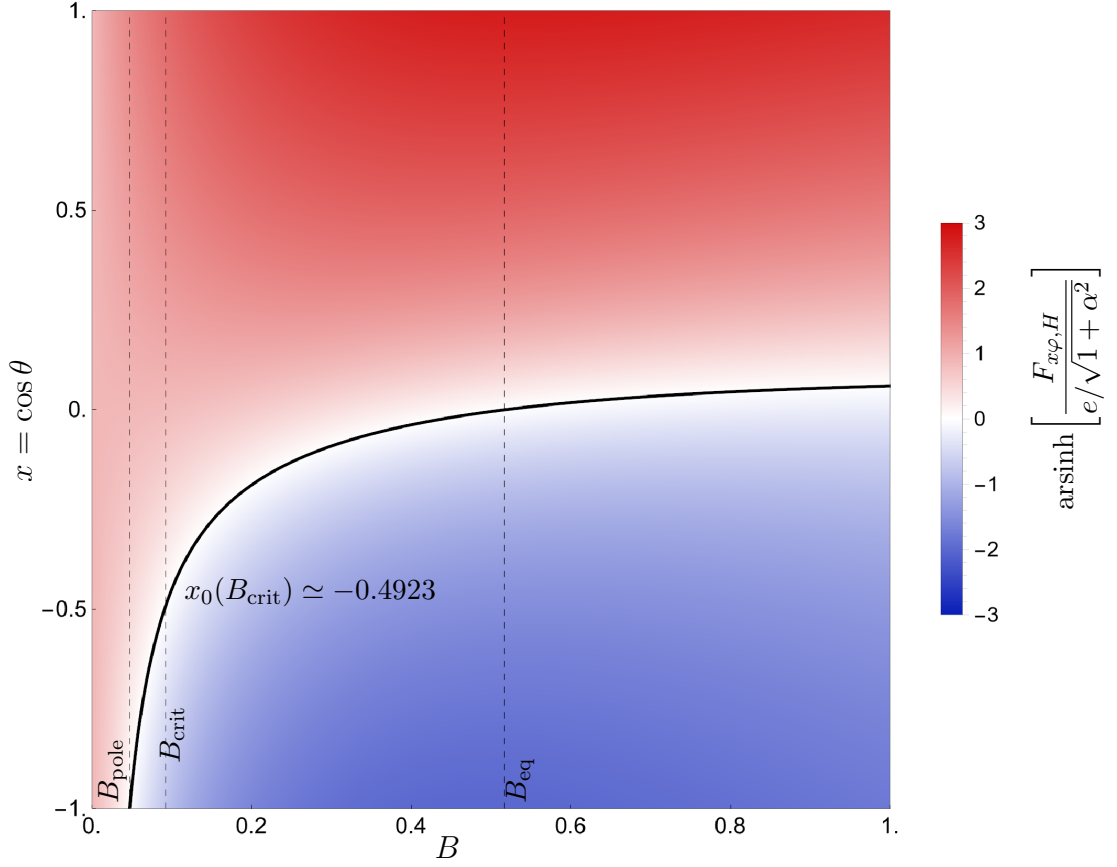
\begin{figure}[H]
\centering
\begin{overpic}[width=0.82\textwidth]{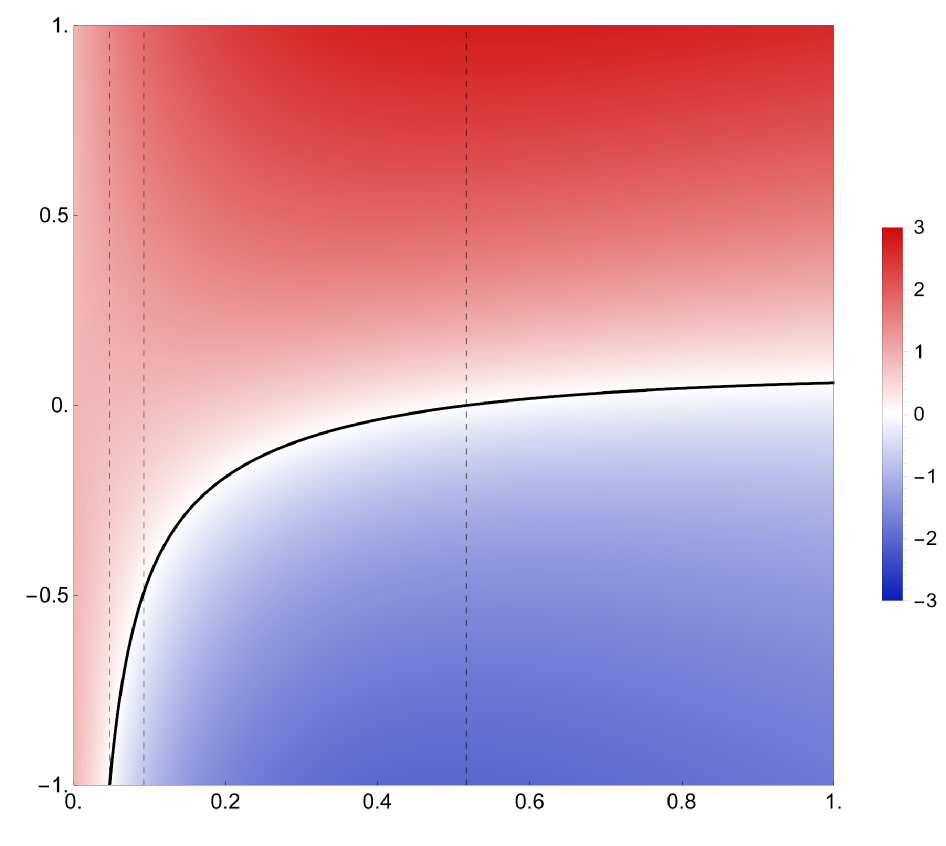}
%
%
\put(43.5,2.5){\makebox(0,0){$B$}}
\put(0.5,45.5){\rotatebox{90}{\makebox(0,0){$x=\cos\theta$}}}
%
\put(8.3,7.0){\rotatebox{90}{\small $B_{\rm pole}$}}
\put(14.5,11.0){\rotatebox{90}{\small $B_{\rm crit}$}}
\put(49.4,7.0){\rotatebox{90}{\small $B_{\rm eq}$}}
%
\put(17.0,27.5){\small $x_0(B_{\rm crit})\simeq-0.4923$}
%
\put(106.0,49.0){\rotatebox{90}{\makebox(0,0){%
\small
$\displaystyle
\operatorname{arsinh}\!\left[
\frac{F_{x\varphi,H}}{e/\sqrt{1+\alpha^2}}
\right]$}}}
\end{overpic}
\caption{Continuous local magnetic-field reversal structure for
\((m,e,\alpha)=(2,7/10,4/5)\).  Red and blue denote positive and negative
\(F_{x\varphi,H}\), respectively, while the heavy black curve is the local
reversal boundary \(F_{x\varphi,H}=0\).  The three dashed lines mark, from
left to right, \(B_{\rm pole}\), \(B_{\rm crit}\), and \(B_{\rm eq}\).
A reversed south-polar cap first appears at \(B_{\rm pole}\).  At
\(B_{\rm crit}\) the integrated southern flux vanishes although the local zero
remains at \(x_0\simeq-0.4923\); the zero reaches the equator only at
\(B_{\rm eq}\).  The \(\operatorname{arsinh}\) transformation compresses the
displayed dynamic range without changing either the sign or the reversal
locus.}
\label{magnetic density fig}
\end{figure}

Figure~\ref{magnetic density fig} gives a continuous view of the same reversal
process in the \((B,x)\) plane. Unlike Fig.~\ref{horizon flux redistribution fig},
it displays the normalized coordinate component $F_{x\varphi,H}$ rather
than the fractional change in normal flux density. Its sign and zeros
nevertheless coincide with those of ${\cal B}_H$, since ${\cal H}_H>0$.
The inverse-hyperbolic sine is used only to compress
the color dynamic range and therefore does not alter either the sign or the zero of
\(F_{x\varphi,H}\).  The heavy black curve is the physical reversal locus
\(F_{x\varphi,H}=0\); the three vertical dashed lines mark
\(B_{\rm pole}\), \(B_{\rm crit}\), and \(B_{\rm eq}\), respectively.

At fixed $0<e<m$, Eq.~\eqref{exact reversal latitude} gives the
strong-field limit
\begin{align}
 x_\infty
 =\lim_{B\to\infty}x_0(B)
 =\frac{m-\sqrt{m\sigma}}{e},
 \qquad 0<x_\infty<1.
 \label{strong field reversal latitude}
\end{align}
The upper bound follows from
$(m-\sqrt{m\sigma})^2< m^2-\sigma^2=e^2$ for $0<\sigma<m$.
Thus every fixed nonextremal charged member retains a finite northern cap
with the original field orientation in the strong-field limit.
For $(m,e)=(2,7/10)$, the exact expression gives
$x_\infty\simeq0.09183$.
This fixed-$\sigma$ strong-field regime should be distinguished from the
correlated near-extremal strong-field limit of
Sec.~\ref{near extremal magnetic reversal sec}, in which $\sigma\to0$ and
$B\to\infty$ together; the two limits do not commute, and the latter
instead drives the reversal boundary to the north pole.

\subsection{Intrinsic deformation and conical curvature}
\label{intrinsic horizon geometry sec}

For the north-regular choice $C=C_N$ and $Be>0$, the residual conical
deficit lies on the south axis. We now examine the horizon dilaton and
intrinsic curvature, and compare their north--south variation with the
conical and magnetic responses.

\subsubsection{Horizon dilaton and Gaussian curvature}

Intrinsic horizon curvature and Euclidean embedding diagrams have long been useful diagnostics of distorted black-hole horizons; sufficiently strong distortions can generate negative Gaussian-curvature regions and obstruct a global embedding in Euclidean three-space~\cite{FrolovShoom2007}.  We use the same diagnostics here to isolate the deformation produced by the BR environment and the residual conical structure.

The near-horizon series gives
\begin{align}
 {
 K_H(x)=
 \frac{2\sigma}{m+\sigma}
 \frac{
 1+B^2m^2-Be\sqrt{1+B^2m^2}\,x
 }{
 1+B^2\sigma^2x^2
 }}.
 \label{KH horizon}
\end{align}
Using the matter-field relation in Eq.~\eqref{final matter},
\begin{align}
 {
 \phi_H(x)=
 -\frac{\alpha}{1+\alpha^2}\ln K_H(x)}.
 \label{horizon dilaton}
\end{align}
The north--south difference is
\begin{align}
 {
 \phi_N-\phi_S=
 \frac{\alpha}{1+\alpha^2}
 \ln\!\left[
 \frac{1+B^2m^2+Be\sqrt{1+B^2m^2}}
      {1+B^2m^2-Be\sqrt{1+B^2m^2}}
 \right]}.
 \label{dilaton NS}
\end{align}
For $B\neq0$ the horizon scalar is generally latitude dependent.  Its north--south
difference vanishes when $Be=0$.

For the canonical horizon metric in Eq.~\eqref{horizon canonical metric}, the Gaussian
curvature takes the compact form
\begin{align}
 K_G(x)=-\frac{F''(x)}{2{\cal H}_H}.
 \label{Gaussian curvature}
\end{align}
It is therefore convenient to define
\begin{align}
 \widehat K(x)\equiv{\cal H}_H K_G(x)=-\frac12F''(x).
\end{align}
In the weak-background expansion,
\begin{align}
 \widehat K(x)
 =
 1+\frac{6Be}{1+\alpha^2}\,x
 +B^2K_2(x)+O(B^3),
 \label{weak curvature}
\end{align}
where $K_2(x)$ is even in $x$. Pole values of the smooth curvature denote
limits from the regular part of the horizon and exclude the distributional
conical contribution. The leading north--south curvature asymmetry is
\begin{align}
 \widehat K_N-\widehat K_S
 =
 \frac{12Be}{1+\alpha^2}+O(B^3).
 \label{weak curvature NS}
\end{align}
For $e=0$, the exact horizon geometry is invariant under $x\to-x$.

The exact expression can yield negative smooth curvature for sufficiently
strong background fields. For example, numerical evaluation at
\begin{equation*}
 (m,e,\alpha,B)=(2,7/10,4/5,2)
\end{equation*}
gives
\begin{align}
 \widehat K(0)\simeq-0.482904,
 \qquad K_G(0)\simeq-0.225662.
\end{align}
The zeros bounding the negative-curvature region containing the equator
are $x\simeq-0.163307$ and $x\simeq0.188225$. This example lies beyond
the range $0\leq B\leq1$ used for the embedding checks below.

The coupling
$\alpha=1$ is special in this respect.  In this case the horizon function reduces to
\begin{align}
 F_1(x)
 =
 N_1\,\frac{1-x^2}{D_0-D_1x},
\end{align}
where
\begin{align}
 D_0=1+B^2m^2,\qquad
 D_1=Be\sqrt{1+B^2m^2},\qquad
 N_1=\sqrt{1+B^2\sigma^2}(1+B^2m^2)^{1/4}.
\end{align}
Its dimensionless curvature is
\begin{align}
 \widehat K_1(x)
 =
 N_1\,\frac{D_0^2-D_1^2}{(D_0-D_1x)^3}.
 \label{alpha1 curvature}
\end{align}
Since
\begin{align}
 D_0^2-D_1^2=(1+B^2m^2)(1+B^2\sigma^2)>0
\end{align}
and $D_0>|D_1|$, Eq.~\eqref{alpha1 curvature} is positive throughout the nonextremal
horizon.

Combining Eq.~\eqref{weak curvature NS} with the weak-field tension
in Eq.~\eqref{mu weak B} gives
\begin{align}
 \widehat K_N-\widehat K_S
 =
 12\mu_S+O(B^2).
 \label{curvature tension relation}
\end{align}
The combination $Be/(1+\alpha^2)$ controls both the conical tension
and the north--south curvature difference at linear order. Beyond this
order they are distinct measures of the black-hole/background interaction.

\subsubsection{Isometric embeddings}

The conical normalization must be included when considering a Euclidean
surface-of-revolution embedding.  Introducing a $2\pi$-periodic angle
$\psi=\varphi/C$, the embedding functions obey
\begin{align}
 R_C(x)&=C\sqrt{{\cal H}_H F(x)},\\
 \frac{Z_C'(x)^2}{{\cal H}_H}
 &=
 \frac{4-C^2F'(x)^2}{4F(x)}.
 \label{embedding condition}
\end{align}
Thus a local Euclidean surface-of-revolution embedding requires $4-C^2F'(x)^2\geq0$ and depends on which
azimuthal normalization, $C=C_N$ or $C=C_S$, is chosen.  In the special uncharged
$\alpha=1$ sector, $C_N=C_S=(1+B^2m^2)^{1/4}$ and the regular horizon embeds as the
round sphere
\begin{align}
 R^2+Z^2=R_0^2,\qquad
 R_0=\frac{2m}{(1+B^2m^2)^{1/4}},
 \label{alpha1 round sphere}
\end{align}
consistent with
$K_G=R_0^{-2}=\sqrt{1+B^2m^2}/(4m^2)$.

We consider a family at fixed $(m,e,\alpha)$, choosing the azimuthal
period so that the north exterior axis remains regular as $B$ varies:
\begin{align}
 C=C_N(B).
\end{align}
The value of $C_N(B)$ varies along this family; it is the choice of
regular axis, rather than the numerical azimuthal period, that is fixed.
The local metric is specified by ${\cal H}_H$ and $F(x)$.  A representative example,
\begin{align}
 (m,e,\alpha)=\left(2,\frac{7}{10},\frac45\right),
\end{align}
was studied for increasing $B$.  The north-regular completion satisfies the
Euclidean surface-of-revolution condition throughout the horizon for all sampled
values $0\leq B\leq1$.

\begin{figure}[t]
\centering
\begin{minipage}[t]{0.325\textwidth}
\centering
\textbf{(a)}\quad \(B=0\)

\vspace{-2mm}
\includegraphics[width=1.05\linewidth,trim=0 85 0 0,clip]{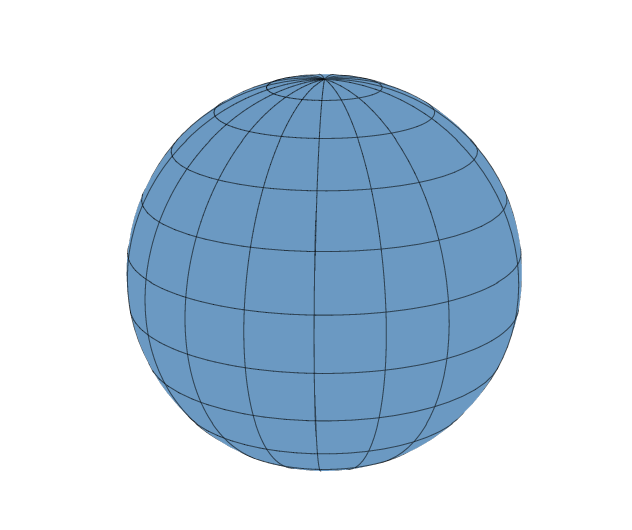}%
\par\vspace{3mm}
\textit{Isolated horizon}

{\small Spherical reference geometry.}
\end{minipage}
\hfill
\begin{minipage}[t]{0.325\textwidth}
\centering
\textbf{(b)}\quad \(B=3/25\)

\vspace{-2mm}
\includegraphics[width=1.05\linewidth,trim=0 85 0 0,clip]{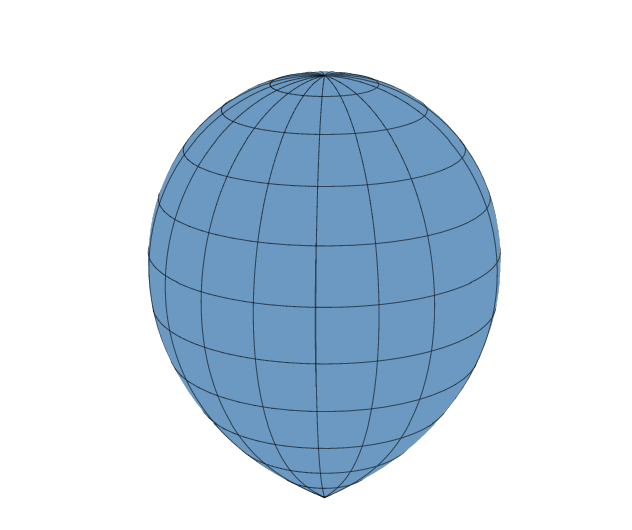}%
\par\vspace{3mm}
\textit{Moderate BR field}

{\small Azimuthal contraction and north--south asymmetry.}
\end{minipage}
\hfill
\begin{minipage}[t]{0.325\textwidth}
\centering
\textbf{(c)}\quad \(B=1/2\)

\vspace{-2mm}
\includegraphics[width=1.05\linewidth,trim=0 85 0 0,clip]{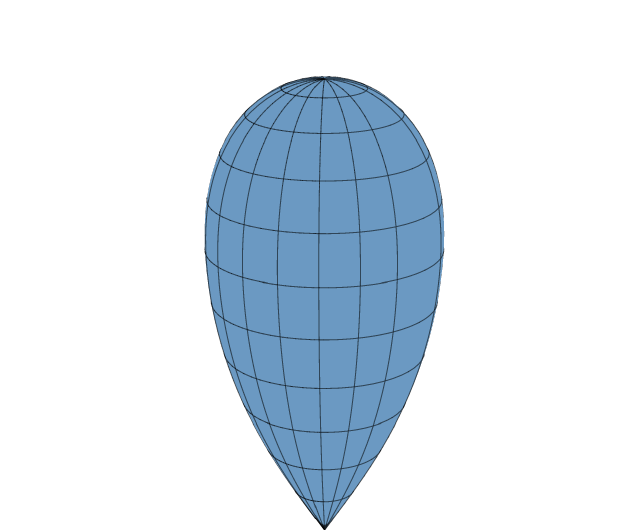}%
\par\vspace{3mm}
\textit{Strong BR field}

{\small Pronounced contraction and south-polar conical deformation.}
\end{minipage}

\caption{Three-dimensional isometric embeddings of the north-regular horizon,
\(C=C_N\), for \((m,e,\alpha)=(2,7/10,4/5)\), shown on a common physical
scale.  Increasing the BR field reduces the circumferential scale and
produces an increasingly pronounced north--south asymmetry.  The north pole
is regular by construction, whereas the pointed south pole represents the
residual conical singularity.}
\label{horizon embedding 3d fig}
\end{figure}

Figure~\ref{horizon embedding 3d fig} shows that, along the sampled
north-regular family, increasing $B$ reduces the circumferential
scale and strengthens the north--south asymmetry. The increasingly
pointed south pole reflects the growing conical deficit measured
by $\mu_S$. The embeddings therefore display both smooth intrinsic
deformation and the residual conical singularity. The curvature
profiles discussed below describe the regular part of the horizon;
the distributional contribution at the south pole is evaluated
separately through the Gauss--Bonnet relation.

\begin{samepage}
For comparison, at fixed $B$ the alternative choice $C=C_S$ changes only the
azimuthal normalization, not the local functions $F(x)$ and $K_G(x)$.
For the positive-$Be$ examples above, the south-regular completion fails the
ordinary Euclidean surface-of-revolution condition near the opposite north pole,
whereas the sampled north-regular horizons admit such embeddings across
the regular part, terminating at the conical south pole.  Under
$B\to-B$ together with $x\to-x$, the two situations are interchanged:
\begin{align}
 F(x;B)=F(-x;-B),\qquad C_N(B)=C_S(-B).
 \label{embedding reflection}
\end{align}
Thus reversing the external field reflects the horizon geometry north--south.
\end{samepage}

The same trend can be characterized intrinsically, without reference to the
embedding coordinate $Z_C$.  Writing $x=\cos\theta$ and
$F(x)=(1-x^2)G(x)$ removes the integrable coordinate singularities at the poles,
so that
\begin{align}
 ds_{\rm mer}
 =
 \sqrt{\frac{{\cal H}_H}{G(\cos\theta)}}\,d\theta .
\end{align}
Define the north and south pole-to-equator meridional lengths
\begin{align}
 L_N&=\int_0^{\pi/2}
 \sqrt{\frac{{\cal H}_H}{G(\cos\theta)}}\,d\theta,\\
 L_S&=\int_{\pi/2}^{\pi}
 \sqrt{\frac{{\cal H}_H}{G(\cos\theta)}}\,d\theta,
\end{align}
and
\begin{align}
 q_{\rm mer}=\frac{L_N-L_S}{L_N+L_S}.
 \label{q mer}
\end{align}
\begin{samepage}
For the same three members, corresponding to $B=0$, $3/25$, and $1/2$,
respectively, the resulting values are
\begin{align}
 q_{\rm mer}=0,\qquad -0.032,\qquad -0.096 .
 \label{q mer examples}
\end{align}
\end{samepage}
For these three examples, the magnitude of the meridional asymmetry
increases with $B$. The negative values indicate that the south
pole-to-equator meridian is longer than its northern counterpart. The common
factor $\sqrt{{\cal H}_H}$ cancels from $q_{\rm mer}$, which therefore measures
relative asymmetry independently of the overall horizon scale. It is also
independent of the azimuthal period $C$. The embeddings combine this
north--south difference in meridional length with a decrease in equatorial
circumferential size.

\begin{figure}[H]
\centering
\begin{overpic}[width=0.74\textwidth]{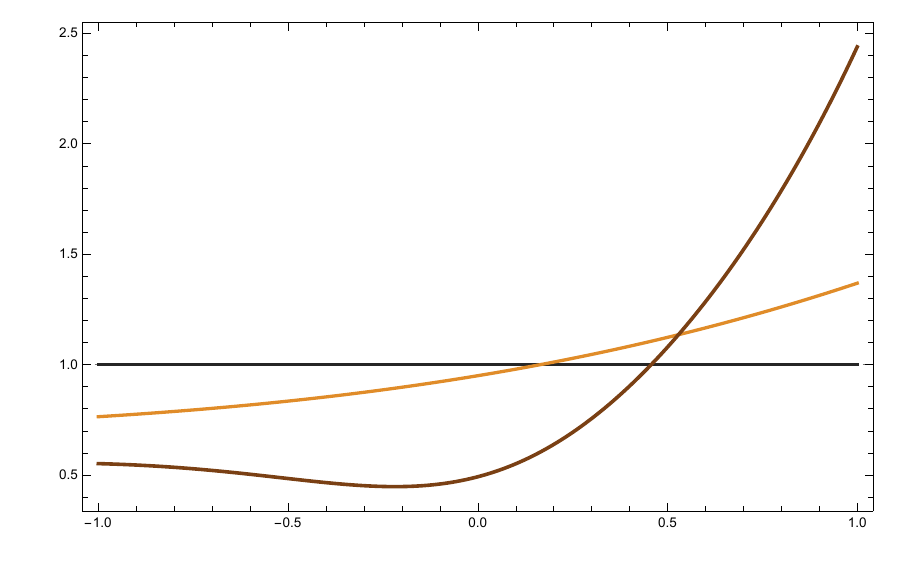}
\put(51,-3){\makebox(0,0){$x=\cos\theta$}}
\put(-3,32){\rotatebox{90}{\makebox(0,0){$\widehat K(x)$}}}
\end{overpic}
\par\vspace{5mm}
\caption{Intrinsic Gaussian-curvature redistribution for
\((m,e,\alpha)=(2,7/10,4/5)\).  The dimensionless profile
\(\widehat K(x)\equiv{\cal H}_H K_G(x)=-F''(x)/2\) is shown against
\(x=\cos\theta\) for \(B=0\), \(3/25\), and \(1/2\).  The dashed horizontal
line marks the round-horizon value \(\widehat K=1\).  The black, orange, and brown curves correspond respectively to $B=0$, $3/25$, and $1/2$;
the $B=0$ curve coincides with the dashed reference line.  Increasing the BR field
produces a north--south redistribution of intrinsic curvature, complementing
the deformation visible in the three-dimensional embeddings.}
\label{horizon curvature profiles fig}
\end{figure}

The accompanying smooth intrinsic deformation can be characterized independently by the
dimensionless Gaussian-curvature profile
\(\widehat K(x)={\cal H}_H K_G(x)=-F''(x)/2\) defined above.
Figure~\ref{horizon curvature profiles fig} shows this intrinsic curvature
redistribution for the same three members.  The \(B=0\) curve gives
\(\widehat K=1\) everywhere, while increasing \(B\) produces a progressively
stronger north--south asymmetry.

\subsubsection{Distributional curvature and Gauss--Bonnet}

\label{gauss bonnet conical sec}

The conical singularity also has a direct intrinsic interpretation through
the Gauss--Bonnet theorem. Let $\mathcal H_{\rm reg}$ denote the horizon
cross-section with its two poles removed. For the constant-determinant
horizon metric
\begin{align}
 d\ell_H^2={\cal H}_H\left[\frac{dx^2}{F(x)}+F(x)d\varphi^2\right],
 \qquad 0\leq\varphi<2\pi C,
\end{align}
the smooth Gaussian curvature and area element satisfy
\begin{align}
 K_G=-\frac{F''(x)}{2{\cal H}_H},
 \qquad
 dA={\cal H}_H\,dx\,d\varphi .
\end{align}
The smooth-curvature integral is therefore determined entirely by the two pole
derivatives,
\begin{align}
 I_{\rm smooth}(C)
 \equiv\int_{\mathcal H_{\rm reg}}K_G\,dA
 =-\pi C\left[F'(1)-F'(-1)\right].
 \label{smooth curvature integral}
\end{align}
For an arbitrary azimuthal normalization the two conical contributions are
\begin{align}
 \delta_N(C)&=2\pi\left(1-\frac{C}{C_N}\right),&
 \delta_S(C)&=2\pi\left(1-\frac{C}{C_S}\right).
\end{align}
Using $C_N=-2/F'(1)$ and $C_S=2/F'(-1)$ gives the exact identity
\begin{align}
 I_{\rm smooth}(C)+\delta_N(C)+\delta_S(C)=4\pi .
 \label{gauss bonnet conical identity}
\end{align}
Thus the total curvature required by the Gauss--Bonnet theorem remains $4\pi$,
but it need not be carried entirely by the smooth Gaussian-curvature profile.

For the north-regular completion $C=C_N$, $\delta_N=0$ and
$\delta_S=8\pi\mu_S$, so Eq.~\eqref{gauss bonnet conical identity} reduces to
\begin{align}
 \int_{\mathcal H_{\rm reg}}K_G\,dA
 =4\pi-\delta_S
 =4\pi-8\pi\mu_S .
 \label{smooth curvature tension relation}
\end{align}
Thus $8\pi\mu_S$ is the integrated distributional Gaussian-curvature
contribution at the residual south pole.  The function $K_G(x)$ plotted above accounts only for the smooth curvature
over the regular part of the horizon.

Along the north-regular family with $e>0$, increasing $B$ redistributes
the smooth intrinsic curvature and increases the fraction of the total
Gauss--Bonnet integral carried by the south-pole deficit. For the
representative family $(m,e,\alpha)=(2,7/10,4/5)$, the smooth fraction
$I_{\rm smooth}/(4\pi)$ decreases from $1$ at $B=0$ to approximately $0.770$ at
$B=1/2$ and $0.705$ at $B=10$, while the complementary conical fractions are
approximately $0.230$ and $0.295$, respectively.  Their sum remains identically
unity. Equation~\eqref{smooth curvature tension relation} therefore identifies the same conical tension
in the exterior-axis regularity condition and the horizon Gauss--Bonnet integral.

\section{Near-extremal limit and strong-field scaling}
\label{extremal sec}

We approach the extremal endpoint through the nonextremal family
to examine the geometric regularity of the limiting horizon and
the fate of the magnetic-reversal hierarchy. We hold
$(m,B,\alpha,C)$ fixed and take $\sigma\to0^+$, rather than
imposing $e=m$ directly on the full solution. Unlike the
north-regular families illustrated above, this prescription
holds the numerical azimuthal normalization $C$ fixed.
We parameterize this limit by
\begin{align}
 \sigma=\sqrt{m^2-e^2}\longrightarrow0^+,
 \qquad
 e=\sqrt{m^2-\sigma^2},
 \label{controlled extremal path}
\end{align}

\subsection{Area and temperature scaling}

The horizon area density found above has the near-extremal form
\begin{align}
 {\cal H}_H
 &\sim h_0\,
 \sigma^{\frac{2\alpha^2}{1+\alpha^2}},
 \label{extremal HH scaling}\\
 h_0
 &=
 4^{\frac{\alpha^2}{1+\alpha^2}}
 m^{\frac{2}{1+\alpha^2}}
 (1+B^2m^2)^{-\frac{1}{2(1+\alpha^2)}} .
\end{align}
Thus the Einstein--Maxwell case $\alpha=0$ retains a finite horizon area,
whereas every $\alpha>0$ member of the dilatonic family has
\begin{align}
 A_H\longrightarrow0 .
\end{align}
The BR field changes the coefficient of this collapse but not its scaling exponent.

The temperature, however, need not follow the area in the same way.  For the surface gravity
it is useful to employ the exact nonextremal identity
\begin{align}
 \kappa{\cal H}_H=\sigma,
 \qquad
 \kappa A_H=4\pi C\,\sigma ,
 \label{area kappa identity}
\end{align}
which also fixes the normalization of the near-extremal temperature.  Hence
\begin{align}
 \kappa
 &\sim
 \kappa_0\,
 \sigma^{\frac{1-\alpha^2}{1+\alpha^2}},
 \label{extremal kappa scaling}\\
 \kappa_0
 &=
 4^{-\frac{\alpha^2}{1+\alpha^2}}
 m^{-\frac{2}{1+\alpha^2}}
 (1+B^2m^2)^{\frac{1}{2(1+\alpha^2)}} .
\end{align}
For the dilatonic family with \(\alpha>0\), the limiting behavior separates
into three regimes:
\begin{align}
 0<\alpha<1 &: \quad A_H\to0,\qquad T_H\to0,\\
 \alpha=1 &: \quad A_H\to0,\qquad
 T_H\to
 \frac{(1+B^2m^2)^{1/4}}{4\pi m},
 \label{alpha one extremal temperature}\\
 \alpha>1 &: \quad A_H\to0,\qquad T_H\to\infty .
 \label{three extremal regimes}
\end{align}
The condition $\sigma\to0$ should therefore not by itself be interpreted as
producing an ordinary regular zero-temperature extremal horizon.  The value \(\alpha=1\) separates the temperature scaling of the endpoint,
but it does not determine its geometrical regularity.  That question requires the scalar
and curvature diagnostics below.

\begin{figure}[H]
\centering
\begin{minipage}[t]{0.48\textwidth}
\centering
\textbf{(a)}\\[-1mm]
\begin{overpic}[width=\linewidth]{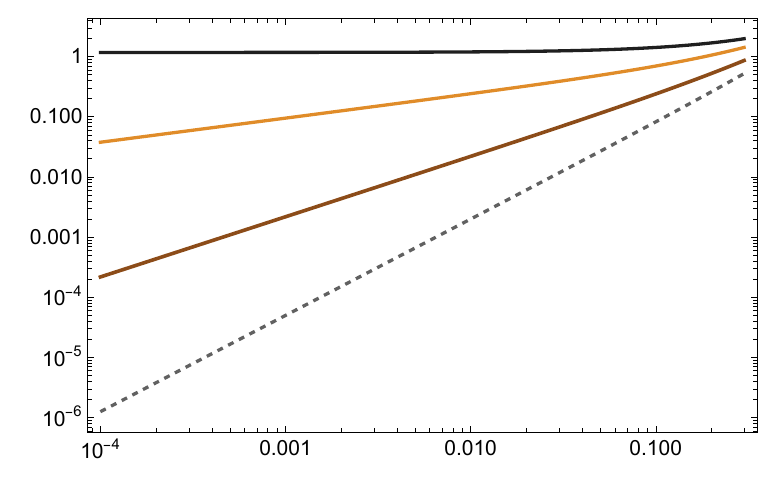}
\put(50,2){\makebox(0,0){\normalsize$\sigma/m$}}
\put(-2,50){\rotatebox{90}{\makebox(0,0){\normalsize$\mathcal H_H/m^2$}}}
\end{overpic}
\end{minipage}
\hfill
\begin{minipage}[t]{0.48\textwidth}
\centering
\textbf{(b)}\\[-1mm]
\begin{overpic}[width=\linewidth]{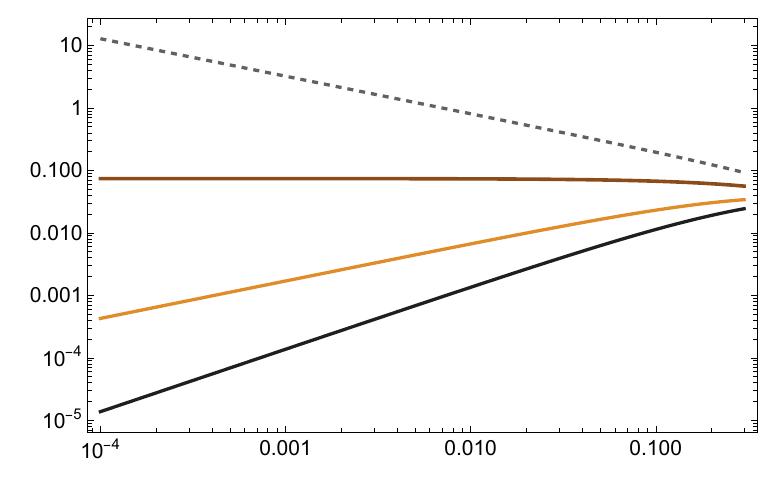}
\put(50,2){\makebox(0,0){\normalsize$\sigma/m$}}
\put(-2,50){\rotatebox{90}{\makebox(0,0){\normalsize$mT_H$}}}
\end{overpic}
\end{minipage}

\caption{Controlled near-extremal scaling at fixed $m$ and $B$ for representative
dilaton couplings.  Panel (a) shows the dimensionless horizon area density
$\mathcal H_H/m^2$, which remains finite in the Einstein--Maxwell case $\alpha=0$ but vanishes
for every $\alpha>0$ member as $\sigma\to0^+$.  Panel (b) shows the
dimensionless Hawking temperature $mT_H$, which tends to zero for $0\leq\alpha<1$,
approaches a finite nonzero limit at $\alpha=1$, and diverges for $\alpha>1$.
The black, gold, brown and gray dashed curves correspond respectively to
$\alpha=0$, $1/2$, $1$ and $2$.
The curves use the exact nonextremal expressions along
$e=\sqrt{m^2-\sigma^2}$; the asymptotic slopes agree with the exponents
derived above.}
\label{extremal scaling fig}
\end{figure}

Figure~\ref{extremal scaling fig} summarizes these near-extremal area and temperature
scalings.

The same fixed-parameter scalings also determine the entropy--temperature relation.
With $S_H=A_H/4$, eliminating $\sigma$ gives, for $0<\alpha<1$,
\begin{align}
 S_H
 \propto
 T_H^{\frac{2\alpha^2}{1-\alpha^2}} .
 \label{low alpha ST scaling}
\end{align}
For $\alpha>1$, the same elimination instead gives
\begin{align}
 S_H
 \propto
 T_H^{-\frac{2\alpha^2}{\alpha^2-1}} .
 \label{high alpha ST scaling}
\end{align}
The string coupling $\alpha=1$ must be treated separately and cannot be obtained by
continuing either exponent.  There $S_H$ vanishes linearly with $\sigma$ while
$T_H$ approaches the finite value in Eq.~\eqref{alpha one extremal temperature}; the
limiting entropy--temperature slope is
\begin{align}
 \left.\frac{dS_H}{dT_H}\right|_{\rm ext}
 =
 -\frac{8\pi^2 C m^3}{\sqrt{1+B^2m^2}} .
\end{align}
These relations are fixed-parameter near-extremal scaling laws.  Since the BR
background is not asymptotically flat, we do not infer from them a global ADM first law
or an asymptotically-flat heat capacity.

\subsection{Dilaton behavior and curvature singularity}

The temperature behavior does not decide the first question posed above: whether the
limiting horizon is geometrically regular.  We therefore turn to intrinsic and
four-dimensional curvature diagnostics.

The exact horizon scalar is
\begin{align}
 \phi_H(x)
 =
 -\frac{\alpha}{1+\alpha^2}\ln K_H(x),
\end{align}
with $K_H(x)$ given in Eq.~\eqref{KH horizon}.  Along
Eq.~\eqref{controlled extremal path},
\begin{align}
 \phi_H(x)
 =
 -\frac{\alpha}{1+\alpha^2}\ln\sigma
 +\phi_0(x)+O(\sigma),
 \label{extremal phi decomposition}
\end{align}
where
\begin{align}
 \phi_0(x)
 =
 -\frac{\alpha}{1+\alpha^2}
 \ln\left[
 \frac{2}{m}
 \left(
 1+B^2m^2-Bm\sqrt{1+B^2m^2}\,x
 \right)
 \right].
 \label{extremal phi finite}
\end{align}
Thus the divergent part is independent of $x$, while the BR
environment leaves a finite north--south angular profile.  Differentiating the exact
horizon expression before taking the limit gives
\begin{align}
 \lim_{\sigma\to0^+}\partial_x\phi_H
 =
 \frac{\alpha Bm\sqrt{1+B^2m^2}}
 {(1+\alpha^2)
 \left[
 1+B^2m^2-Bm\sqrt{1+B^2m^2}\,x
 \right]} .
 \label{extremal phi derivative}
\end{align}

The intrinsic Gaussian curvature gives the first direct regularity diagnostic.  We
distinguish this two-dimensional curvature of the horizon cross-section from
four-dimensional curvature invariants of the spacetime, which are examined separately
below.
Writing
\begin{align}
 \widehat K\equiv{\cal H}_H K_G=-\frac12F''(x),
\end{align}
one finds a finite limiting dimensionless curvature profile $\widehat K_{\rm ext}(x)$, while
\begin{align}
 K_G(x)
 \sim
 k_{G0}(x)\,
 \sigma^{-\frac{2\alpha^2}{1+\alpha^2}}
 \label{extremal KG scaling}
\end{align}
at generic latitudes.  Hence for every $\alpha>0$ the limiting horizon
cross-section is intrinsically curvature singular, even in the range
$0<\alpha<1$ where the temperature tends to zero.  The Einstein--Maxwell
case $\alpha=0$ is qualitatively different: its area and smooth intrinsic
curvature remain finite. This statement concerns the regular part of the
horizon and does not remove the residual conical singularity.

The singularity is not merely a pathology of the two-dimensional horizon
metric.  In the present normalization the trace of the EMD Einstein equation is
\begin{align}
 R=2(\nabla\phi)^2.
\end{align}
The angular scalar-gradient contribution at the horizon is
\begin{align}
 R_x
 =
 2g^{xx}(\partial_x\phi_H)^2
 =
 \frac{2F}{{\cal H}_H}(\partial_x\phi_H)^2 .
\end{align}
For $\alpha B\neq0$ and $-1<x<1$, Eq.~\eqref{extremal phi derivative} is
finite and generically nonzero, and therefore
\begin{align}
 R_x
 \sim
 {\cal R}_0(x)\,
 \sigma^{-\frac{2\alpha^2}{1+\alpha^2}},
 \qquad {\cal R}_0(x)>0 .
 \label{extremal Ricci scaling}
\end{align}
For each fixed $\sigma>0$, the radial scalar derivative is finite at interior
horizon latitudes and $g^{rr}$ vanishes there. Thus the radial contribution
to $R$ vanishes in the horizon limit, and $R_H=R_x$. For $\alpha B\ne0$,
the four-dimensional Ricci scalar therefore diverges at generic interior
horizon latitudes along the near-extremal family.
The vanishing of this particular angular contribution at the symmetry-axis
endpoints does not alter the conclusion that the limiting horizon is singular.

For $\alpha>0$ and nonzero $B$, the BR field modifies the angular profiles
and coefficients of the near-extremal laws, while the characteristic powers at generic
interior latitudes are
\begin{align}
 {\cal H}_H&\sim\sigma^{2p},&
 K_G,\ R_H&\sim\sigma^{-2p},
 \qquad p=\frac{\alpha^2}{1+\alpha^2}.
\end{align}
The value $\alpha=1$ separates the temperature regimes, whereas $\alpha=0$
separates finite smooth horizon curvature from the divergent curvature of
the dilatonic family. The Einstein--Maxwell endpoint still carries the
residual conical singularity when $Be\ne0$.

The zero-background case requires a distinction between limiting procedures.
For $B=0$, the scalar depends only on $r$, and the Ricci scalar is
\begin{align}
 R=\frac{2\alpha^2r_-^2}{(1+\alpha^2)^2r^4}
 \left(1-\frac{r_+}{r}\right)
 \left(1-\frac{r_-}{r}\right)^{-\frac{1+3\alpha^2}{1+\alpha^2}}.
 \label{isolated Ricci scalar}
\end{align}
It vanishes on every nonextremal horizon. However, taking $r_+=r_-=m$
in the exterior expression before approaching $r=m$ gives
\begin{align}
 R_{\rm ext}
 =\frac{2\alpha^2m^2}{(1+\alpha^2)^2r^4}
 \left(1-\frac{m}{r}\right)^{-\frac{2\alpha^2}{1+\alpha^2}},
 \label{isolated extremal Ricci scalar}
\end{align}
which diverges as $r\to m^+$ for every $\alpha>0$. Thus the isolated
extremal solution is also curvature singular, although this is not detected
by first evaluating $R$ on each nonextremal horizon. The intrinsic horizon
curvature already diverges along that family. These results are consistent
with the singular extremal limit of the potential-free charged EMD solution
\cite{KalloshPeet1992,GibbonsMaeda1988,GHS1991}.

For every $\alpha>0$, the near-extremal endpoint along the fixed-$B$ path is therefore
singular, irrespective of its temperature scaling. We next examine the
fate of the magnetic-reversal hierarchy.

\subsection{Magnetic reversal and strong-field double scaling}
\label{near extremal magnetic reversal sec}

The magnetic analysis of Sec.~\ref{magnetic redistribution sec} provides the
relevant starting point.  As $\sigma\to0^+$, the intrinsic monopole flux approaches a finite nonzero value
while the horizon collapses, so an increasingly strong BR field is required to overcome the
outward monopole field in the south.  The local normal magnetic-flux density can therefore diverge
while the reversal thresholds simultaneously recede to infinity: magnitude and
orientation are distinct questions.

More precisely, along the controlled fixed-$B$ near-extremal path the local field magnitude behaves as
\begin{align}
 {\cal B}_H(x)
 \sim b_0(x)\,
 \sigma^{-\frac{2\alpha^2}{1+\alpha^2}}
 \label{extremal magnetic field scaling}
\end{align}
at generic latitudes.  Thus for $\alpha>0$ a finite limiting integrated magnetic charge
is supported on a collapsing horizon by a divergent local normal flux density:
the endpoint exhibits local magnetic-flux concentration rather than flux expulsion.
Here the comparison with the black-hole Meissner effect should be understood
carefully: the flux concentrated here is the finite limiting intrinsic monopole flux,
whereas the standard Meissner statement concerns expulsion of externally sourced
horizon-threading flux from a regular extremal horizon~\cite{BicakJanis1985,GibbonsPangPope2014}.
This fixed-$B$ concentration concerns the field magnitude; the reversal thresholds below instead probe its orientation and require a correlated near-extremal strong-field limit.

The three magnetic thresholds have a common near-extremal scaling.  Using the
exact expressions above,
\begin{align}
 B_{\rm pole}
 &\sim\frac{1}{2\sqrt{m\sigma}},&
 B_{\rm crit}
 &\sim\frac{1}{\sqrt{2m\sigma}},&
 B_{\rm eq}
 &\sim\frac{1}{\sqrt{m\sigma}},
\end{align}
so that
\begin{align}
 B_{\rm pole}:B_{\rm crit}:B_{\rm eq}
 \longrightarrow
 1:\sqrt2:2 .
 \label{near extremal magnetic hierarchy}
\end{align}
Thus no finite fixed \(B\) reaches a reversal threshold as \(\sigma\to0^+\).
At the same time, their common $\sigma^{-1/2}$ divergence identifies the natural
strong-field zoom under which the reversal structure can remain finite:
\begin{align}
 \sigma\to0^+,\qquad
 B\to\infty,\qquad
 \mathfrak b\equiv B\sqrt{m\sigma}
 \quad\hbox{fixed}.
 \label{magnetic double scaling}
\end{align}
In the correlated scaling~\eqref{magnetic double scaling}, the reduced quadratic in
Eq.~\eqref{reversal quadratic} becomes at leading order
\begin{align}
 \mathcal Q_H(x)\longrightarrow
 1-\mathfrak b^2(1-x)^2.
\end{align}
The root that continuously describes the moving reversal boundary within
$-1\leq x\leq1$ is therefore
\begin{align}
 x_0(\mathfrak b)=1-\frac{1}{\mathfrak b},
 \qquad \mathfrak b\geq\frac12 .
 \label{universal reversal curve}
\end{align}

For $0<\mathfrak b<1/2$, this root lies outside the horizon interval and
there is no local reversal. The other root, $x=1+1/\mathfrak b$, lies
outside the horizon interval. The same limiting law follows directly from
Eq.~\eqref{exact reversal latitude}.

\begin{figure}[H]
\centering
\begin{overpic}[width=0.72\textwidth]{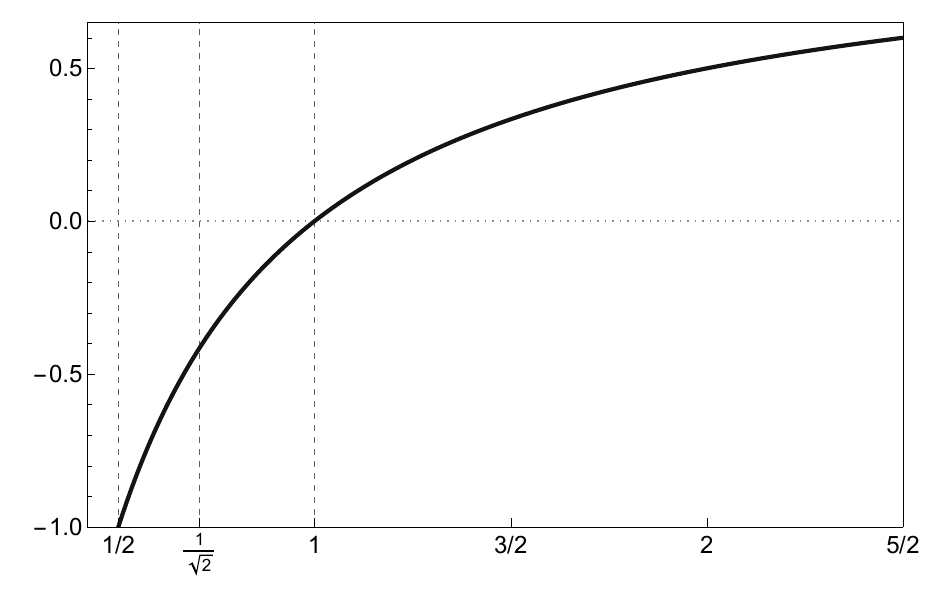}
\put(2,50){\rotatebox{90}{\makebox(0,0){\normalsize$x_0=\cos\theta_0$}}}
\end{overpic}

\vspace{-1mm}
\centering\normalsize$\mathfrak b=B\sqrt{m\sigma}$

\caption{Universal magnetic-reversal law in the correlated near-extremal strong-field
limit $\sigma\to0^+$, $B\to\infty$ with
$\mathfrak b=B\sqrt{m\sigma}$ fixed.  The physical reversal latitude is
$x_0(\mathfrak b)=1-1/\mathfrak b$ for $\mathfrak b\geq1/2$; no physical
reversal occurs below this threshold. The vertical dashed lines mark
$\mathfrak b_{\rm pole}=1/2$, $\mathfrak b_{\rm crit}=1/\sqrt2$, and
$\mathfrak b_{\rm eq}=1$, corresponding respectively to the birth of a reversed
south-polar cap, vanishing integrated southern flux, and passage of the reversal
boundary through the equator.  Thus the thresholds divide the plot into four
regimes: no reversal, local south-polar reversal, hemispheric reversal, and
reversal extending into the northern hemisphere.}
\label{universal reversal fig}
\end{figure}

The exact finite-$\sigma$ reversal branch therefore reduces at leading order to a
one-parameter universal law.  The correlated scaling eliminates the detailed
dependence on $m$, $e$, and $\alpha$ at leading order while preserving the
physical ordering of the reversal process.  The three
transitions become
\begin{align}
 \mathfrak b_{\rm pole}&=\frac12,&
 \mathfrak b_{\rm crit}&=\frac1{\sqrt2},&
 \mathfrak b_{\rm eq}&=1,
\end{align}
with
\begin{align}
 x_0(\mathfrak b_{\rm crit})=1-\sqrt2 .
\end{align}
Figure~\ref{universal reversal fig} summarizes the resulting double-scaled phase
structure.  In contrast with the full finite-$\sigma$ reversal locus of
Fig.~\ref{magnetic density fig}, the near-extremal reversal structure is governed by the single
curve $x_0(\mathfrak b)=1-1/\mathfrak b$ and the three fixed thresholds above.

The finite-$\sigma$ hierarchy
$B_{\rm pole}<B_{\rm crit}<B_{\rm eq}$ therefore survives as
$1/2<1/\sqrt2<1$, but its detailed parameter dependence has disappeared.  The correlated
limit preserves the same sequence of physical events in a universal one-parameter form.

The fixed-nonextremal strong-field limit and the correlated near-extremal
limit describe different asymptotic regimes. At fixed nonextremal \(\sigma\),
\(B\to\infty\) gives the finite latitude \(x_\infty<1\), so a northern cap
retains the original monopole orientation no matter how large \(B\) becomes.
By contrast, the near-extremal double-scaled theory has
\(x_0=1-1/\mathfrak b\), and a subsequent \(\mathfrak b\to\infty\) sends
the reversal boundary to the north pole.  The two procedures therefore describe
different competitions between horizon collapse and external-field
amplification. The subsequent large-$\mathfrak b$ limit refers to the limiting
reversal law and does not assert a uniform approximation at the north pole.

The two questions posed at the start of the section therefore have distinct answers.
Along the fixed-$(m,B,\alpha,C)$ near-extremal path considered here, every $\alpha>0$ member approaches a zero-area, curvature-singular endpoint,
even when $T_H\to0$.  The magnetic-reversal hierarchy, however, survives on the correlated
scale $B\sim\sigma^{-1/2}$ and reduces to the universal law
$x_0=1-1/\mathfrak b$.

\section{Conclusion}
\label{conclusion sec}

We have constructed an explicit static, magnetically charged EMD
solution that generalizes the centered RN--BR black hole for arbitrary
dilaton coupling. In canonical Weyl coordinates, the construction replaces
the Schwarzschild horizon-rod potential by the RN rod potential of half-length
$\sigma=\sqrt{m^2-e^2}$ while retaining the background contribution to the
auxiliary harmonic function. Integrating the nonlinear background--rod
interaction in $\widetilde\gamma_{\rm RNBR}$ determines the remaining metric
factor. The Einstein--Maxwell, uncharged, zero-background and pure-background
limits are recovered with the coordinate and normalization prescriptions
specified above. The resulting solution has also been checked numerically
against the full EMD field equations. For nonzero coupling, the pure-background limit
is the magnetic EMD deformation of BR, rather than the homogeneous
Einstein--Maxwell product geometry.

The horizon identity $\kappa{\cal H}_H=\sigma$ relates the area density and
surface gravity directly to the Weyl rod scale. The latitude dependence is
carried separately by $F(x)$. For $Be\ne0$, the two exterior-axis segments
require different azimuthal periods. Choosing the north axis to be regular
leaves a south-axis tension $\mu_S(Y,\alpha)$, where
$Y=Be/\sqrt{1+B^2m^2}$. For $Be>0$, this tension increases with $B$, remains
bounded at fixed $0<e/m<1$, and is suppressed as $\alpha^2$ increases.

The same interaction redistributes the horizon magnetic flux. The charge
is conserved between enclosing surfaces within each solution, while the
hemispheric polarization satisfies $Z=mY/(m-\sigma)$ for $e\ne0$.
For $0<e<m$ and increasing $B>0$, the ordered thresholds
$B_{\rm pole}<B_{\rm crit}<B_{\rm eq}$ distinguish local south-polar reversal,
vanishing integrated southern flux, and equatorial crossing. These thresholds
and the reversal latitudes are independent of $\alpha$ at fixed seed
parameters $(m,e)$; the local field magnitude and intrinsic geometry are not.

The intrinsic geometry provides a complementary description of the conical
response. Its weak-field curvature asymmetry satisfies
$\widehat K_N-\widehat K_S=12\mu_S+O(B^2)$, while the north-regular embeddings
display both smooth deformation and a conical south pole. The exact relation
\begin{equation*}
 \int_{\mathcal H_{\rm reg}}K_G\,dA=4\pi-8\pi\mu_S
\end{equation*}
identifies $8\pi\mu_S$ as the integrated distributional Gaussian-curvature
contribution at that pole.

At fixed $(m,B,\alpha,C)$, the near-extremal horizon area vanishes for every
$\alpha>0$, and the smooth intrinsic curvature diverges at generic latitudes.
For nonzero fixed $B$, the horizon Ricci scalar also diverges at generic interior
latitudes. At $B=0$, the
extremal exterior geometry is singular although the Ricci scalar vanishes
on each nonextremal horizon. The limiting temperature is zero, finite or
divergent according as $0<\alpha<1$, $\alpha=1$ or $\alpha>1$; temperature
alone therefore does not determine regularity. The finite limiting monopole
flux becomes concentrated on the collapsing horizon. Meanwhile, the common
threshold scaling $B\sim\sigma^{-1/2}$ selects the correlated variable
$\mathfrak b=B\sqrt{m\sigma}$ and the universal reversal law
$x_0=1-1/\mathfrak b$ for $\mathfrak b\geq1/2$.

\medskip

The centered restriction $\ell=\beta_2-\beta_1=0$ can first be relaxed
within the underlying Weyl family~\cite{Astorino2026RNBR}. Restoring a relative
axial displacement would test its effect on the conical imbalance and
horizon polarization, and whether it provides additional freedom toward balance.

The purely magnetic representative used throughout this paper also has an immediate
electric counterpart.  The potential-free EMD equations admit the discrete
electric--magnetic duality
\begin{align}
 \phi\longrightarrow-\phi,
 \qquad
 F\longrightarrow e^{-2\alpha\phi}\,{}^\star F ,
\end{align}
with the Einstein-frame metric unchanged.  Thus the corresponding electrically charged
black hole in the electric EMD deformation of the BR background requires no separate
metric construction.  A dyonic extension requires additional structure.  The
Emparan--Teo transformation used here is adapted to purely electric or purely magnetic
static seeds and does not directly generate independent electric and magnetic sectors~\cite{EmparanTeo2001}.  For generic dilaton coupling the dyonic EMD problem is also substantially less tractable~\cite{PolettiEtAl1995,GibbonsLuPope2023}.  A general dyonic RN--BR solution would therefore require solution-generating structure beyond that employed here.

Stationary and multi-black-hole generalizations provide two further directions.  Exact
Kerr and Kerr--Newman black holes in a BR magnetic background are now known in
Einstein--Maxwell theory~\cite{Ovcharenko:2026KNBR}, but rotation introduces the twist sector and lies outside the
static Weyl construction used here.  Their EMD counterparts would consequently require
a stationary generalization of the present method.

Likewise, recent Einstein--Maxwell work has produced Majumdar--Papapetrou-type multi-black-hole
configurations with BR asymptotics~\cite{Furugori:2026multi}.  From the Weyl viewpoint developed here, a
nonextremal multi-rod extension would naturally replace the single finite-rod potential by
a sum of rod potentials, so that the corresponding first-order Weyl equations would
contain both background--rod and pairwise rod--rod interaction terms.  Whether such configurations admit useful EMD balance
conditions, especially in view of the singular limiting behavior of the purely magnetic dilatonic solution found here,
is an interesting open problem.

A particularly natural and more immediate extension is to introduce acceleration.  The
centered solution studied here is generically conically unbalanced when both the intrinsic
charge and the external BR field are present, suggesting that an additional geometrical
degree of freedom should enter the more general balance problem.  In Einstein--Maxwell
theory accelerating charged black holes in a BR background are already known~\cite{OvcharenkoPodolsky2026}.  It is
therefore natural to investigate whether the EMD construction used here can be extended
to the corresponding accelerating seed.  The resulting spacetime would provide a
BR-background generalization of the dilatonic C-metric of Dowker, Gauntlett, Kastor and
Traschen~\cite{Dowker:1993bt}, recovered when the external BR field is removed.  Earlier
solution-generating constructions of dilatonic C-metrics and black diholes, including
configurations with unbalanced magnetic charges and accelerating diholes, provide useful
precedents for this program~\cite{EmparanTeo2001,Liang:2001ea,Teo:2005cz}.  In particular, acceleration may provide the additional
freedom needed to remove the residual conical singularity of the centered solution.
Work on this accelerating generalization is currently in progress and will be reported
elsewhere.

\end{document}